\documentclass{article}

\usepackage{fullpage}
\usepackage[skip=5pt]{parskip}
\usepackage[utf8]{inputenc}
\usepackage[T1]{fontenc}
\usepackage[english]{babel}
\usepackage{hyperref}
\hypersetup{
    colorlinks=true,
    linkcolor=cyan,
    filecolor=cyan,      
    urlcolor=cyan,
    citecolor=cyan,
    }
    
\usepackage{graphicx}
\usepackage{import}
\usepackage{tikz}
\usetikzlibrary{patterns, arrows.meta, patterns.meta}
\usetikzlibrary{positioning}

\usepackage{stmaryrd}
\usepackage{eurosym} % symbole euro
\usepackage{xcolor}
\usepackage{amsthm}
\usepackage{amsmath}
\usepackage{amssymb}
\usepackage{amsfonts}
\usepackage{mathrsfs}
\usepackage{bbm}

\usepackage{algorithm}
\usepackage{algpseudocode}

\usepackage{mathabx}
\DeclareMathSymbol{\cap}{\mathbin}{symbols}{"5C} % jolis cap's
 
\newcommand{\ind}{\mathbbm{1}}
\newcommand{\BB}{\mathbb{B}}

\newcommand{\EE}{\mathbb{E}}

\newcommand{\LL}{\mathbb{L}}

\newcommand{\NN}{\mathbb{N}}

\newcommand{\PP}{\mathbb{P}}

\newcommand{\RR}{\mathbb{R}}

\newcommand{\VV}{\mathbb{V}}

\DeclareMathOperator*{\argmax}{arg\,max}
\newcommand*{\norm}[1]{\left\lVert#1 \right\rVert}
\renewcommand{\epsilon}{\varepsilon}

\title{\huge\textsc{Macroscopic Activity-Based Modeling}\\
\textsc{of Urban Active Mobility}}
\author{Romain Aza\"{\i}s$^\dagger$, Adrien Marion$^\dagger$, and Florian Patout$^\ddagger$}
\date{\vspace{-0.3cm}\small$^\dagger$\,Inria Lyon, France\qquad$^\ddagger$\,INRAE Avignon, France}

\begin{document}

\maketitle

\vspace{-0.35cm}

\noindent\makebox[\linewidth]{\rule{\textwidth}{0.4pt}}\\[5pt]
\textbf{Abstract.} This paper develops a macroscopic, activity-based model of urban active mobility using nonintrusive sensor data. It introduces attendance functions to describe spatio-temporal travel patterns between activities and formulates the disaggregation of aggregated counts as a statistical inference problem. Counts are modeled as Poisson variables, and unknown subpopulation sizes are estimated via maximum likelihood, with theoretical guarantees and an efficient EM algorithm for computation. Grounded in a microscopic stochastic model, the framework offers a scalable and privacy-preserving approach to analyzing urban soft mobility dynamics.\\[3pt]
\noindent\makebox[\linewidth]{\rule{\textwidth}{0.4pt}}

\vspace{-0.35cm}

\section{Introduction}

Today's cities face rapid growth and pressing environmental challenges. To achieve the transition to more sustainable urban environments, transportation plays a pivotal role, as promoted by the United Nations' 11th Sustainable Development Goal\footnote{\url{https://docs.un.org/en/A/RES/71/313}, last consulted on April 12, 2026}. The effort to characterize and quantify urban transport sustainability -- and to identify the efforts required to achieve it -- is fundamental in shaping and supporting public policy \cite{akande2019lisbon, banister2008sustainable}. A precise understanding of urban mobility dynamics is particularly essential to this effort \cite{barbarossa2020post,gebhardt2016intermodal,liu2025study}. 

In this article, we examine travel patterns in public spaces, specifically for soft mobility modes such as walking and cycling. Although individual trips are inherently unpredictable, collective patterns emerge when journeys between two different activities -- such as shopping, leisure and work -- are grouped together \cite{valee2024mobiliscope}. The aim of this article is to develop a deterministic model that captures the dynamics of each group. It is derived from the large-population limit of a microscopic, individual-based stochastic model. Combined with street-level traffic observations, it offers a precise characterization of expected travel dynamics.

Traditionally, descriptive studies on mobility have relied on surveys \cite{cirillo2004mobilite}. In France,  EMC$^2$ surveys are conducted in various urban areas across the country \cite{rabaud2018emc2}, overseen by Cerema. They provide a snapshot of residents' travel patterns over the course of a typical day at a macroscopic scale (district, city), taking into account the mode of transportation used, the purpose of the trip, age, gender, and other factors. Conducting such surveys is resource-intensive. For example, the ongoing survey in the Lyon metropolitan area is expected to take several years and cost \euro2.4 million\footnote{\url{https://www.francemarches.com/files/BOAMP/25-70604.html}, last consulted on April 12, 2026}. Additionally, investigators face growing challenges in reaching or recruiting households, due to both declining public interest and logistical barriers -- such as the lack of landline connections, intercom systems at building entrances, and other access-related issues \cite{rabaud2018emc2}.

Another class of studies make use of the global system for mobile communications (GSM) data collected by carriers to reconstruct urban travel patterns \cite{bonnel2015passive}. Various methods for analyzing this data have been developed to complement field studies \cite{fekih2019methodologie,xu2015understanding}. However, they raise significant issues, particularly regarding anonymization and selection bias \cite{keusch2023coverage, rodriguez2018detecting}. GPS data constitute an additional source of tracking information. These are more accurate than data from cell towers, but at the cost of stricter anonymization constraints. European studies conducted to date have been carried out, but only on a limited panel of volunteers over periods not exceeding a few months \cite{depeau2014recherche,sila2016analysis}.

In this article, we propose a complementary approach to these existing methods, focusing on automatic urban counts from sensors. This data is nonintrusive, free from selection bias, openly accessible, and updated in real time \cite{livingston2021predicting,yang2010investigating}. It is highly reliable, with minimal gaps in the time-series records. Automatic counters are becoming widespread in major cities, with accelerated deployment in recent years: in Lyon, for example, 184 bike counters and 72 pedestrian counters are now operational, including 40 installed since 2021\footnote{\url{https://data.grandlyon.com/portail/fr/jeux-de-donnees/sites-comptage-metropole-lyon/info}, last consulted on April 12 2026}. These devices capture local traffic counts every hour for the main activity mobility modes (pedestrians, electric scooters, bicycles). However, their main drawback is that these counts aggregate trips without distinguishing their link with activities.

We transform the challenge of disaggregating the count data into a statistical estimation problem by introducing a dynamical model for each type of journey, based on additional geographic data of two types: the spatial distribution of activity locations (distribution of shops, residential areas, parks, \textit{etc.}) and activity schedules (hours of work, frequentation of shops, \textit{etc.} \cite{ganault2025enquetes}). Using these data, we model the expected dynamics for each journey type over the study area and within a fixed time interval by designing trips between each pair of activities. This model captures hourly and location-specific movement trends, represented by an \textit{attendance function}. This function is by definition dimensionless, akin to a frequency function. It provides a spatio-temporal mapping that encapsulates all the travel information for a given journey type. It will be a key insight of our study.   

Each attendance function is associated with a scale parameter. It represents the size of the corresponding subpopulation, \textit{i.e.} the number of trips made across the study area and time interval for the given journey type. Combined together, the attendance function and the vector of subpopulation sizes give an expected number of passages at each location and each time step. To disaggregate the hidden counts, our strategy is thus to infer these scaling parameters, represented as a vector whose dimension corresponds to the number of journey types.

To that end, we develop a statistical framework that links the count data to the previous model. Count data are modeled as a noisy version of the expected number of passages. This yields an analytical expression for the likelihood involving the attendance function and the count data, depending on the unknown vector of subpopulation sizes. This vector is also inferred using the Maximum Likelihood Estimator (MLE). The statistical efficiency of this estimator improves with increasing data volume, and it is possible to quantify the associated uncertainties. In terms of asymptotics, several approaches are possible: reducing the counting time period, concatenating data over multiple days (or other cycle of a repeated phenomenon), or increasing the total number of counters. We will focus on the latter option, as the observation period is inherently constrained by the counting devices themselves. Aggregating data over multiple days would undermine the value of real-time analysis and, moreover, introduces challenges in identifying comparable or homogeneous days.

The maximum likelihood estimator we study does not admit a closed-form solution. To compute it, we adapt the Expectation-Maximization (EM) algorithm, which is particularly well-suited to our context of hidden data. The inputs are the count data and the attendance functions, evaluated at the counter's locations. Its theoretical efficiency is matched by the rapid execution of the expectation and maximization steps in our context. Finally, this model provides precise insight into the dynamics of soft modes of transportation across the entire area for each activity, all while relying solely on non-invasive data. 

In transportation modeling, macroscopic approaches are traditionally based on transport equations, often nonlinear, as in classical vehicular traffic models \cite{lighthill1955kinematic,richards1956shock}. These types of equations are closely related to a fluid mechanics perspective: the flux depends on density and interaction effects, allowing realistic congestion effects to emerge, but at the cost of potentially high theoretical and computational complexity. This partly explains why alternative models have been developed at different scales and levels of approximation, aiming to bridge the gap between microscopic and macroscopic descriptions (see, for instance, \cite{di2017follow}). Pedestrian modeling has followed a different path, exploring how individuals in a crowd react to the movement of others, for instance in an evacuation scenario. A large part of the literature is dedicated to microscopic models, to take into account individual behavior, either incorporating stochasticity, as in \cite{henderson1971statistics}, or in a fully deterministic manner, such as in the celebrated social-force model of \cite{helbing1995social}. We refer to the complete review \cite{korbmacher2023time} on this topic. At the macroscopic level, continuum formulations have been proposed, notably in \cite{hughes2002continuum,hughes2003flow}, and later extended to density constrained PDE models \cite{maury2010macroscopic,maury2009handling} (see also the review \cite{santambrogio2018crowd}). The equations capture congestion and collective effects as limits of underlying microscopic models.

It is important to note that our contribution differs from, while remaining complementary to, the aforementioned literature. At the spatial scales (a neighborhood or an entire city) and temporal scales (from a few hours to a full day) considered here, and for the transportation modes under study (active mobility modes), interactions between individuals are of limited relevance. Our objective is to predict mobility flows by trip purpose at a coarse spatial and temporal resolution, without accounting for congestion phenomena, which are certainly important in other contexts. Moreover, in the Lyon metropolitan area, with a temporal discretization on the order of one hour and fewer than 200 counting stations for approximately 10\,000 road segments, only macroscopic patterns can realistically be captured, while interaction effects are likely to remain unobservable. That is why we focus on a low-interaction regime and propose a model designed to be compatible with public counting data. Our linear macroscopic model is derived from a microscopic description. The resulting dynamics are governed by transport equations with \textit{source} and \textit{death} terms, rather than nonlinear closure laws. This allows a direct link between aggregated sensor data and activity-based mobility patterns, while remaining consistent with classical continuum frameworks adapted to active mobility flows.

The rest of this article is organized as follows. In Section~\ref{Counters_and_(des)agregation} we link counting data to attendance functions and to the unknown vector of subpopulation sizes. Together, these objects allow us to construct expected passages in every point and at every time, for each type of journey, that should be close to the hidden data. As the expected passages depend on the unknown vector of subpopulation sizes, Section~\ref{statistical_model} introduces the statistical tools to estimate this parameter using the maximum likelihood method, and presents asymptotic results on its accuracy. In Section~\ref{section_em} we propose an adaptation of the EM algorithm to compute the MLE, addressing the computational challenges of the estimation process. Finally we provide a precise way to construct and interpret the attendance functions in the unidimensional case, by studying a microscopic, individual-based stochastic model in large population. Throughout the article, we illustrate our results using a guiding example, whose expert static data and count data are generated \textit{in silico}. Appendix~\ref{appendix_stat} provides detailed explanations of the classical results for the MLE, presenting the classical results that we adapt to fit our specific framework. Appendix~\ref{appendix_em} includes the computations for the expectation and maximisation steps of our algorithm presented in Section~\ref{section_em}. Finally Appendix \ref{appendix_attendance} contains proofs of formulas presented in Section~\ref{stochastic_deterministic}.

\section{Setting of the problem}\label{Counters_and_(des)agregation}

Our strategy to study travel dynamics is based on data from automatic, non intrusive, counters.
As they only capture localized snapshots that are aggregated across all journey types, we need a method to disaggregate them and link the counts to activities. We introduce the mathematical tools and notations for our model. In this section, we describe the outputs of the counters and clarify the hidden data.

The urban area is a compact metric set $\mathcal{E}$, such as a segment, a subset of $\RR^2$, or a graph, that may represent a street, a field, or a portion of a map. Trips are of $K$ distinct types, each representing a journey between two different activities (leisure, shopping, work, staying home, \textit{etc.}). For instance, commuting trips are treated as two distinct types of journeys: home-to-work and work-to-home; the activities involved are in both ``being at home'' and ``being at work''.

We assume that all individuals are moving without turning back during their trip. This ensures that they are counted at most once by a counter. Inside a fixed time-interval, the counters record the number of individuals passing through their position $x \in \mathcal{E}$. The time step of their data is rough, commonly one hour; we will use the index $i$ for it. Therefore, the number of passages recorder over the time $i$ by the counter at position $x$ is denoted by $n(i, x)$. Additionally, we define the hidden data $n_k(i, x)$ as the number of passages associated with journey type $k$ at the same counter. This hidden data is not directly accessible to this counter, which only records the aggregate sum of all passages,
\begin{equation}\label{aggregated_sum}
n(i, x) = \sum_{k=1}^K n_k(i, x).
\end{equation}
We next show how to to link this count and the associated unknown vector of subpopulation sizes to an attendance function that we define accordingly. This transforms the disaggregation challenge into a problem of estimating the vector of population sizes.

\subsection{From a disaggregation challenge to a vector estimation problem}\label{subsec disagregation vector estimation}

We introduce the notation $N$ for the total number of trips undertaken during the time period of interest $i\leq I$ and within the spatial domain $\mathcal{E}$. A trip is here defined as the path an individual takes from an origin activity to a destination activity.

As we have hidden data, we need to categorize the total number per activity, \textit{i.e.} $N = \left(N_k \right)_{k \leq K}$, where each $N_k$ corresponds to journey of type $k$. For instance, if the time step is one hour and $I = 24$, $N$ is the vector whose coordinate $N_k$ represents the total number of trips corresponding to the journey type $k$, throughout the day and across the entire study area $\mathcal{E}$. 

It should be noted that there is no straightforward relationship linking the count data $n(i,x)$ to the vector $N$: individuals may be counted by multiple counters during their journeys. Furthermore, it is not guaranteed that all individuals are counted at least once by any counter. 

With the introduction of $N$, we can thus now interpret $n_k(i, x)$, the  count of passages of type $k$, as a sample of all the trips traveled over $\mathcal{E}$, of this type $k$. Of course, the difficulty is that $N_k$ is unknown: it should be regarded as oracle data, inaccessible through any field surveys. Furthermore, $n_k(i,x)$ is not directly accessible from the counter. 

To solve these issues, we focus on the \textit{empirical attendance}, defined as the ratio
\begin{equation}\label{empirical_attendance}
    A_k(i, x) := \frac{n_k(i, x)}{N_k}.
\end{equation}
This dimensionless function accounts for  the empirical probability that a trip of type $k$ is observed at location $x$ and time $i$. We can link it to the number of passages recorded with the relationship \eqref{aggregated_sum},
\begin{equation*} % \label{relation_hidden_data}
    n(i, x) = \sum_{k=1}^K N_k A_k(i, x) = \big\langle N, A(i, x) \big \rangle .
\end{equation*}

The key element of the approach is that that the empirical ratio $A_k(i,x)$ is in fact a random perturbation of a \textit{theoretical attendance} $a_k(i, x)$. This function corresponds to patterns in travel behavior between activities. It describes the frequency of visits to a location at given time. For instance, if we consider $k=$~home-to-work commutes, the ratio $a_k(i,x)$ will be significantly higher at $i=$~8 a.m., when many people are traveling to work, compared to $i=$~6 p.m., which is more characteristic of the end of the workday rather than the start. Conversely, for $k=$~work-to-home commutes, the ratio $a_{k}(i,x)$ is expected to be much higher at $i=$~6 p.m. than at $i=$~8 a.m. In other words, $a_k(i, x)$ can be  interpreted as the theoretical probability that an individual undertaking the journey type $k$ will pass through location $x$ during time $i$. In Section~\ref{stochastic_deterministic}, we show that the theoretical attendance can be obtained from a microscopic, individual-based stochastic model based on additional geographic data (spatial distribution of activity locations and activity schedules).

Finally, if we have $N_k$ travels related to the activity $k$, we expect 
\begin{equation}\label{eq hidden data attendance}
    n_k(i, x) \approx N_k a_k(i, x)
\end{equation} 
passages during time interval $i$ through $x$ for the activity $k$. Across all activities, the linear relationship  \eqref{aggregated_sum} means that we further expect $\langle N, a(i, x) \rangle$ passages at the same time interval and position.

\begin{figure}[h]
	\centering
	\def\ptg{1.4}
\def\ptm{0.9}
\def\ptd{0.3}
\def\ptf{0.05}

\def\pdg{0.1}
\def\pdm{0.2}
\def\pdd{1.1}
\def\pdf{1.2}

\def\pug{0.4}
\def\pum{1.2}
\def\pud{0.2}
\def\puf{0.5}

\def\thu{1.5}
\def\thd{1.2}
\def\tht{0.6}

\def\scalecount{.9}
\def\scalewidth{0.7}
\def\swd{1.2}
	
\definecolor{cthu}{rgb}{0.41568627,0.65098039,0.69019608}
\colorlet{cthu_darker}{cthu!75!black}
\definecolor{cthd}{rgb}{0.9372549,0.62745098,0.62745098}
\colorlet{cthd_darker}{cthd!90!black}
\definecolor{ctht}{rgb}{0.627, 0.729, 0.529}
\colorlet{ctht_darker}{ctht!90!black}

%%%%%%%%%%%%%%%%%%%%%%%%%%%%%%%%%%

	\begin{tikzpicture}[scale=1, transform shape]
		\begin{scope}
			\node[anchor=north] at (0.5*12+0.5*5*\scalewidth,3) {Characteristics of journey types: attendance functions and subpopulation sizes};
			
			\node[anchor=south] at (2*\scalewidth, \ptg+.2) {$a_1(\cdot, x)$};
			\draw[pattern={Dots[radius=0.3mm, angle=45, distance=1mm]}] (0,0) rectangle (1*\scalewidth,\pug);
			\draw[pattern={Dots[radius=0.3mm, angle=45, distance=1mm]}] (1*\scalewidth,0) rectangle (2*\scalewidth,\pum);
			\draw[pattern={Dots[radius=0.3mm, angle=45, distance=1mm]}] (2*\scalewidth,0) rectangle (3*\scalewidth,\pud);
			\draw[pattern={Dots[radius=0.3mm, angle=45, distance=1mm]}] (3*\scalewidth,0) rectangle (4*\scalewidth,\puf);
			\node at (.5*\scalewidth, -.3) {$1$};
			\node at (1.5*\scalewidth, -.3) {$2$};
			\node at (2.5*\scalewidth, -.3) {$3$};
			\node at (3.5*\scalewidth, -.3) {$4$};
			\node[anchor=north] at (2*\scalewidth, -.4) {time steps};
			\draw[fill = cthu] (4.6*\scalewidth,0) rectangle (5*\scalewidth, \thu);
			\node[anchor=south] at (4.8*\scalewidth, \ptg+.2) {$N_1$};
		\end{scope}
		
		\begin{scope}[xshift=6cm]
			\node[anchor=south] at (2*\scalewidth, \ptg+.2) {$a_2(\cdot, x)$};
			\draw[pattern={Lines[angle=45, distance=1mm, line width=0.3mm]}] (0,0) rectangle (1*\scalewidth,\pdg);
			\draw[pattern={Lines[angle=45, distance=1mm, line width=0.3mm]}] (1*\scalewidth,0) rectangle (2*\scalewidth,\pdm);
			\draw[pattern={Lines[angle=45, distance=1mm, line width=0.3mm]}] (2*\scalewidth,0) rectangle (3*\scalewidth,\pdd);
			\draw[pattern={Lines[angle=45, distance=1mm, line width=0.3mm]}] (3*\scalewidth,0) rectangle (4*\scalewidth,\pdf);
			\node at (.5*\scalewidth, -.3) {$1$};
			\node at (1.5*\scalewidth, -.3) {$2$};
			\node at (2.5*\scalewidth, -.3) {$3$};
			\node at (3.5*\scalewidth, -.3) {$4$};
			\node[anchor=north] at (2*\scalewidth, -.4) {time steps};
			\draw[fill = cthd] (4.6*\scalewidth,0) rectangle (5*\scalewidth, \thd);
			\node[anchor=south] at (4.8*\scalewidth, \ptg+.2) {$N_2$};
		\end{scope}
		
		\begin{scope}[xshift=12cm]
			\node[anchor=south] at (2*\scalewidth, \ptg+.2) {$a_3(\cdot, x)$};
			\draw[pattern={Lines[angle=0, distance=1mm, line width=0.3mm]}] (0,0) rectangle (1*\scalewidth,\ptg);
			\draw[pattern={Lines[angle=0, distance=1mm, line width=0.3mm]}] (1*\scalewidth,0) rectangle (2*\scalewidth,\ptm);
			\draw[pattern={Lines[angle=0, distance=1mm, line width=0.3mm]}] (2*\scalewidth,0) rectangle (3*\scalewidth,\ptd);
			\draw[pattern={Lines[angle=0, distance=1mm, line width=0.3mm]}] (3*\scalewidth,0) rectangle (4*\scalewidth,\ptf);
			\node at (.5*\scalewidth, -.3) {$1$};
			\node at (1.5*\scalewidth, -.3) {$2$};
			\node at (2.5*\scalewidth, -.3) {$3$};
			\node at (3.5*\scalewidth, -.3) {$4$};
			\node[anchor=north] at (2*\scalewidth, -.4) {time steps};
			\draw[fill = ctht] (4.6*\scalewidth,0) rectangle (5*\scalewidth, \tht);
			\node[anchor=south] at (4.8*\scalewidth, \ptg+.2) {$N_3$};
		\end{scope}
	\end{tikzpicture}
	
%%%%%%%%%%%%%%%%%%%%%%%%%%%%%%%%%%
\vspace{0.5cm}
	
	\begin{tikzpicture}[scale=1, transform shape]
		\node[anchor=north] at (4.75,3.4) {Number of passages at location $x$: expected (and how it is decomposed) \textit{vs.} data};
		
		\begin{scope}[xshift=-\swd*1cm]
			\node[anchor=west] at (-.13, .18+\scalecount*\thu*\pug + \scalecount*\thd*\pdg + \scalecount*\tht*\ptg) {\footnotesize expected};
			\node[anchor=east]at (\swd*1.8, .25+\scalecount*\thu*\pug + \scalecount*\thd*\pdg + \scalecount*\tht*\ptg) {\footnotesize data};
			\draw[pattern={Dots[radius=0.3mm, angle=45, distance=1mm]}, pattern color=cthu_darker]
			(0,0) rectangle (\swd*1, \scalecount*\thu*\pug);
			\draw[pattern={Lines[angle=45, distance=1mm, line width = 0.5mm]}, pattern color = cthd_darker]
			(0,\scalecount*\thu*\pug) rectangle (\swd*1, \scalecount*\thu*\pug + \scalecount*\thd*\pdg);
			\draw[pattern={Lines[angle=0, distance=1mm, line width=0.5mm]}, pattern color=ctht_darker]
			(0,\scalecount*\thu*\pug + \scalecount*\thd*\pdg) rectangle (\swd*1, \scalecount*\thu*\pug + \scalecount*\thd*\pdg + \scalecount*\tht*\ptg);
			\draw[fill = gray!10] (\swd*1,0) rectangle (\swd*2, \scalecount*\thu*\pug + \scalecount*\thd*\pdg + \scalecount*\tht*\ptg+0.1);
			\node[rotate=90] at (\swd*1.5, .2*\scalecount*\thu*\pug + .5*\scalecount*\thd*\pdg + .5*\scalecount*\tht*\ptg+0.2) {$n(\scriptstyle 1 \textstyle, x)$};
			\node at (\swd*1, -.3) {$1$};
		\end{scope}
		
		\begin{scope}[xshift=\swd*1.666cm]
			\node[anchor=west] at (-.13, .18+\scalecount*\thu*\pum + \scalecount*\thd*\pdm+ \scalecount*\tht*\ptm) {\footnotesize expected};
			\node[anchor=east]at (\swd*1.8,.05+\scalecount*\thu*\pum + \scalecount*\thd*\pdm + \scalecount*\tht*\ptm) {\footnotesize data};
			\draw[pattern={Dots[radius=0.3mm, angle=45, distance=1mm]}, pattern color=cthu_darker]
			(0,0) rectangle (\swd*1, \scalecount*\thu*\pum);
			\draw[pattern={Lines[angle=45, distance=1mm, line width = 0.5mm]}, pattern color = cthd_darker]
			(0,\scalecount*\thu*\pum) rectangle (\swd*1, \scalecount*\thu*\pum + \scalecount*\thd*\pdm);
			\draw[pattern={Lines[angle=0, distance=1mm, line width=0.5mm]}, pattern color=ctht_darker]
			(0,\scalecount*\thu*\pum + \scalecount*\thd*\pdm) rectangle (\swd*1, \scalecount*\thu*\pum + \scalecount*\thd*\pdm + \scalecount*\tht*\ptm);
			\draw[fill=gray!10] (\swd*1, 0) rectangle (\swd*2, \scalecount*\thu*\pum + \scalecount*\thd*\pdm + \scalecount*\tht*\ptm -.1);
			\node[rotate=90] at (\swd*1.5, .5*\scalecount*\thu*\pum + .5*\scalecount*\thd*\pdm + .5*\scalecount*\tht*\ptm) {$n(\scriptstyle{2} \textstyle, x)$};
			\node at (\swd*1, -.3) {$2$};
		\end{scope}
		
		\begin{scope}[xshift=\swd*4.333cm]
			\node[anchor=west] at (-.13, .18+\scalecount*\thu*\pud + \scalecount*\thd*\pdd+ \scalecount*\tht*\ptd) {\footnotesize expected};
			\node[anchor=east]at (\swd*1.8,.17+\scalecount*\thu*\pud + \scalecount*\thd*\pdd + \scalecount*\tht*\ptd) {\footnotesize data};
			
			\draw[pattern={Dots[radius=0.3mm, angle=45, distance=1mm]}, pattern color=cthu_darker]
			(0,0) rectangle (\swd*1, \scalecount*\thu*\pud);
			\draw[pattern={Lines[angle=45, distance=1mm, line width = 0.5mm]}, pattern color = cthd_darker]
			(0,\scalecount*\thu*\pud) rectangle (\swd*1, \scalecount*\thu*\pud + \scalecount*\thd*\pdd);
			\draw[pattern={Lines[angle=0, distance=1mm, line width=0.5mm]}, pattern color=ctht_darker]
			(0,\scalecount*\thu*\pud + \scalecount*\thd*\pdd) rectangle (\swd*1, \scalecount*\thu*\pud + \scalecount*\thd*\pdd + \scalecount*\tht*\ptd);
			\draw[fill=gray!10] (\swd*1, 0) rectangle (\swd*2,\scalecount*\thu*\pud + \scalecount*\thd*\pdd + \scalecount*\tht*\ptd + 0.02);
			\node[rotate=90] at (\swd*1.5, .5*\scalecount*\thu*\pud + .5*\scalecount*\thd*\pdd + .5*\scalecount*\tht*\ptd) {$n(\scriptstyle{3} \textstyle, x)$};
			\node at (\swd*1, -.3) {$3$};
		\end{scope}
		
		\begin{scope}[xshift=\swd*7cm]				
			\node[anchor=west] at (-.13, .18+\scalecount*\thu*\puf + \scalecount*\thd*\pdf+ \scalecount*\tht*\ptf) {\footnotesize expected};
			\node[anchor=east]at (\swd*1.8,.19+\scalecount*\thu*\puf + \scalecount*\thd*\pdf + \scalecount*\tht*\ptf) {\footnotesize data};
			
			\draw[pattern={Dots[radius=0.3mm, angle=45, distance=1mm]}, pattern color=cthu_darker]
			(0,0) rectangle (\swd*1, \scalecount*\thu*\puf);
			\draw[pattern={Lines[angle=45, distance=1mm, line width = 0.5mm]}, pattern color = cthd_darker]
			(0,\scalecount*\thu*\puf) rectangle (\swd*1, \scalecount*\thu*\puf + \scalecount*\thd*\pdf);
			\draw[pattern={Lines[angle=0, distance=1mm, line width=0.5mm]}, pattern color=ctht_darker]
			(0,\scalecount*\thu*\puf + \scalecount*\thd*\pdf) rectangle (\swd*1, \scalecount*\thu*\puf + \scalecount*\thd*\pdf + \scalecount*\tht*\ptf);
			\draw[fill=gray!10] (\swd*1, 0) rectangle (\swd*2,\scalecount*\thu*\puf + \scalecount*\thd*\pdf + \scalecount*\tht*\ptf + 0.03);
			\node[rotate=90] at (\swd*1.5, .5*\scalecount*\thu*\puf + .5*\scalecount*\thd*\pdf + .5*\scalecount*\tht*\ptf) {$n(\scriptstyle{4} \textstyle, x)$};
			\node at (\swd*1, -.3) {$4$};
		\end{scope}
		
		\node[anchor=north] at (\swd*4,-.4) {time steps};
			
	\end{tikzpicture}
	
	\caption{Illustration of the disaggregation challenge with three types of journey ($k = 1, 2, 3$) and four time steps ($i = 1, 2, 3, 4$). On the top, the histograms show theoretical attendance functions (at some counter location $x$) and subpopulation sizes for the three journey types. The bottom diagram shows the scalar product at location $x$ for each time interval $i$ between $(a_k(i, x))_{k\leq 3}$ and $(N_k)_{k\leq 3}$, alongside the observed count data $n(i, x)$ (gray rectangle). The scalar products represent the expected number of passages. Thus, they should closely align with the observed counts $n(i,x)$. Since $N$ is unknown, our strategy is to estimate it by determining the best fit between this expected number of passages and the observed counting data.}
	\label{tikz_figure}
\end{figure}
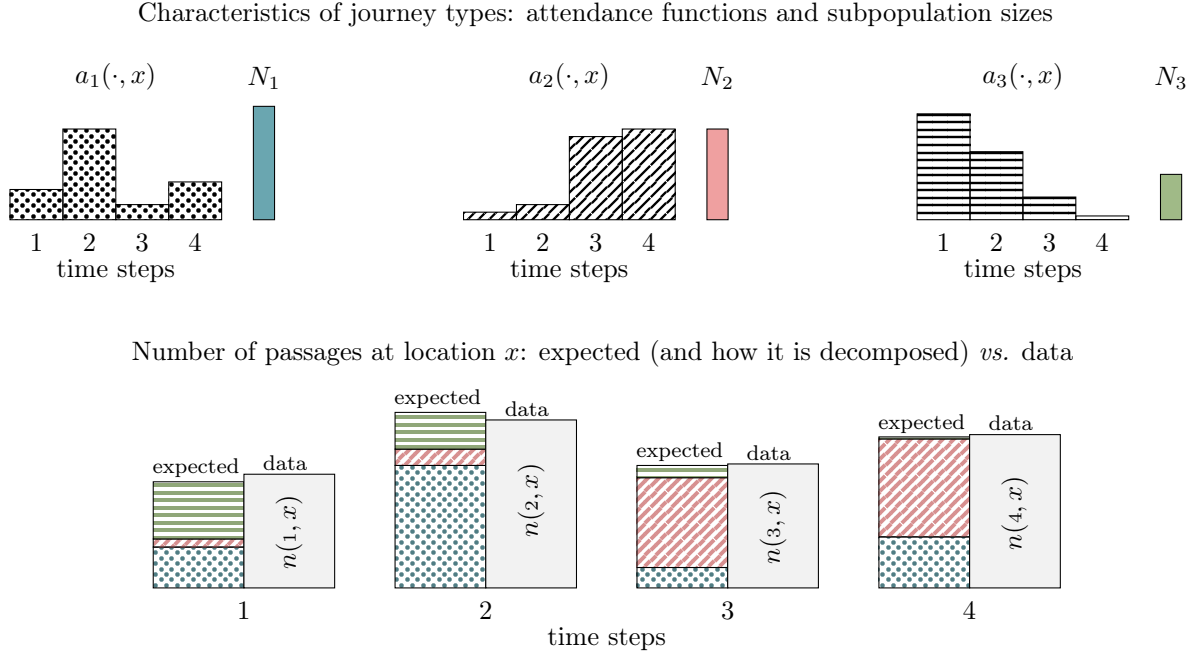

Our first goal in this article is to disaggregate the hourly traffic counts $n$ by travel purpose, using the approximation
\begin{align*}
    n(i, x) \approx \langle N, a(i, x) \rangle.
\end{align*}
We seek to estimate $\widehat{N} \in \RR_+^K$ which aligns the expected number of trips with the observed data transmitted by the counters. We provide a visual explanation of this challenge in Figure~\ref{tikz_figure}.

\subsection{Guiding example: definition}\label{guiding_example_intro}

The results of the paper will be illustrated through the following complete guiding example, generated \textit{in silico}. To define it, we require the two main quantities introduced in the paper, namely a theoretical attendance function $a_k(i,x)$ and counting data $n(i,x)$. To validate our approach, we must demonstrate that our algorithms correctly estimate the scaling parameters $N = (N_k)_{k\leq K}$, which must therefore be known within the framework of this example.

As previously announced, we will explain in Section~\ref{stochastic_deterministic} how to derive the theoretical attendance function from additional geographical data. However, this is not the approach adopted to construct the guiding example. Instead, we assume here that we have access to a simulator (which will be derived in Subsection~\ref{ss:example:theory}) generating counting data $n_k(i,x)$ for each activity -- data that are usually hidden. The theoretical attendance functions are then obtained by averaging (over multiple days) the empirical attendances defined in \eqref{empirical_attendance} from these hidden data. 

\begin{figure}[th]
\centering
\includegraphics[width = .8\textwidth]{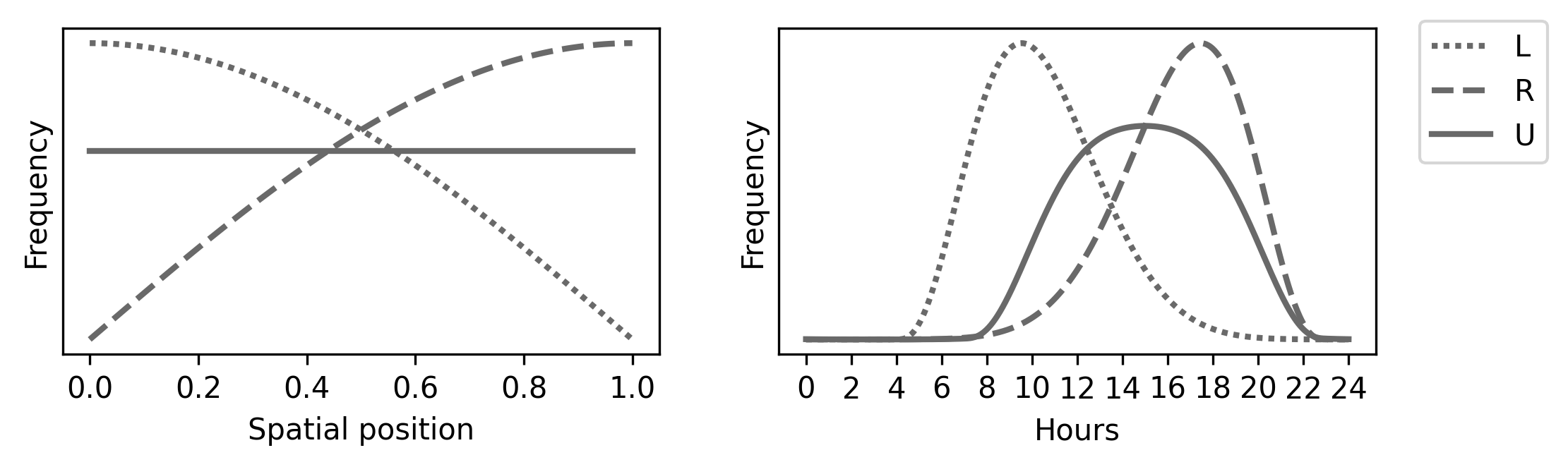}
\caption{Spatial distribution of activity locations (left) and distribution of end time of each activity (right).}
\label{spatial_distribution_agenda}
\end{figure}

We hereby assume that individuals can engage in three activities named $R$, $L$ and $U$ (this can be for example staying at home, shopping and working). The spatial distribution of the activities is not homogeneous: people doing $L$ are predominantly located on the left of the spatial domain, people doing $R$ rather on the right and those doing $U$ are uniformly distributed. Activities are conducted following temporal schedules: the end times of the activities are known (for example the end of work schedule). Spatial distribution and temporal schedule of activities chosen to define the simulator are in Figure~\ref{spatial_distribution_agenda}. Once the theoretical framework for defining attendance $a_k(i,x)$ will be established in Section~\ref{stochastic_deterministic}, the simulator used here to generate the hidden data will be derived in Subsection~\ref{ss:example:theory} from these activity data.

\begin{figure}[th]
    \centering
    \includegraphics[width=1\textwidth]{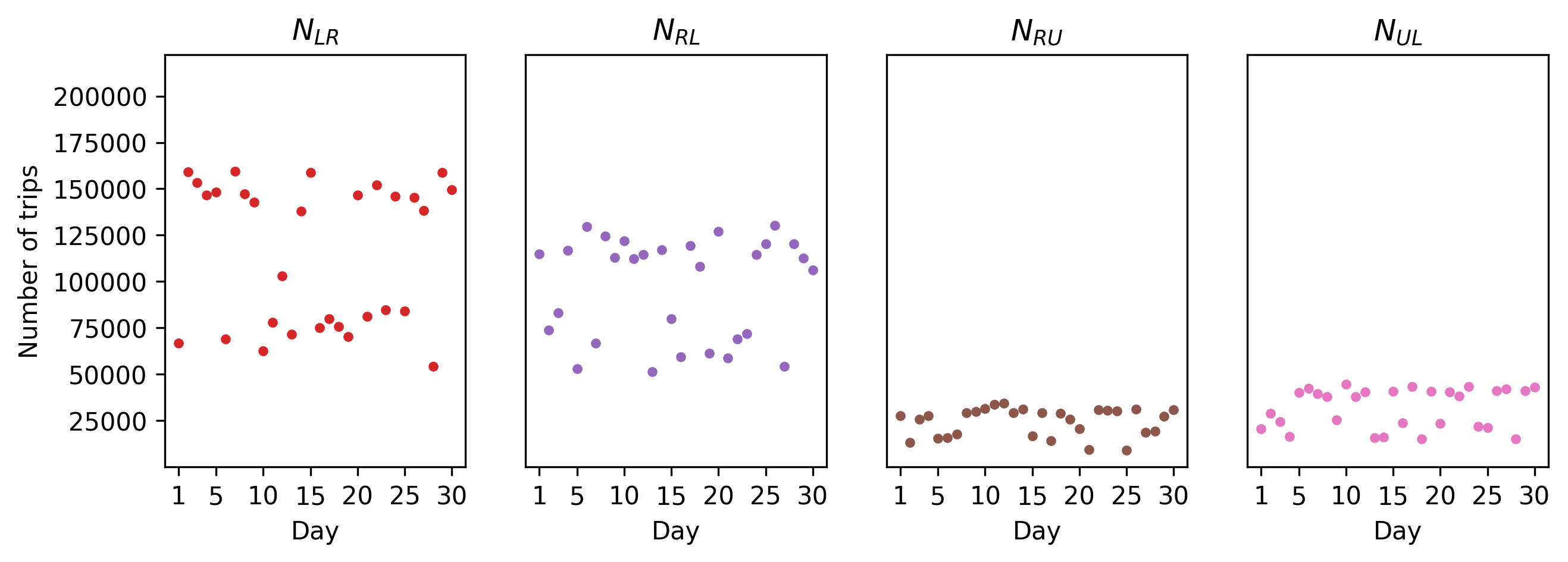}
    \caption{Generated subpopulation sizes for the first $30$ days (among $300$). It can be observed that, for example, $LR$ and $RL$ are much more usual journey types than $RU$ and $UL$.}
    \label{fig ex nbtrips}
\end{figure}

\begin{figure}[th]
\centering
\includegraphics[width = 1\textwidth]{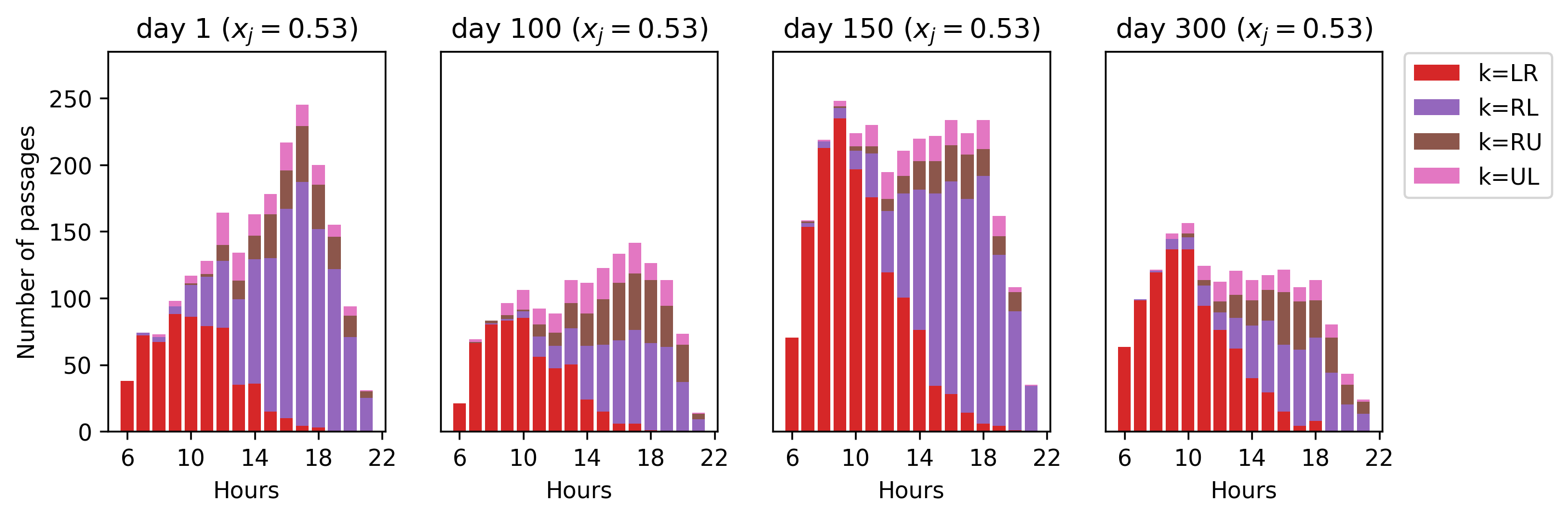}
\caption{Hidden data for different days, at a single counter location $x_j=0.53$ and for each hour of the day. The hidden data are stacked by type of journey.}
\label{fig ex hidden data}
\end{figure}

For clarity reasons, we only  consider four possible types of journey: $LR$ (from $L$ to $R$), $RL$, $RU$ and $UL$. The other types of journey ($LU$ and $UR$) are assumed to be never observed. We generate data \textit{in silico} for $300$ days, which will be used to build the attendance function. More precisely, at day $d\leq 300$, we generate $N_k^{(d)}$ trips for a given type of journey $k \in \{LR, RL, RU, UL \}$ (see Figure~\ref{fig ex nbtrips}). We also randomly place $50$ counters in the space interval $[0,1]$, following a uniform distribution. As aforementioned, the generation of trips then includes the associated path and the type of journey. We have therefore access to the hidden data $n_k^{(d)}(i, x_j)$ at each time $i \leq 24$ and position of each counter $(x_j)_{j \leq 50}$, as illustrated in Figure~\ref{fig ex hidden data}.

\begin{figure}[th]
\centering
    \includegraphics[width=1\textwidth]{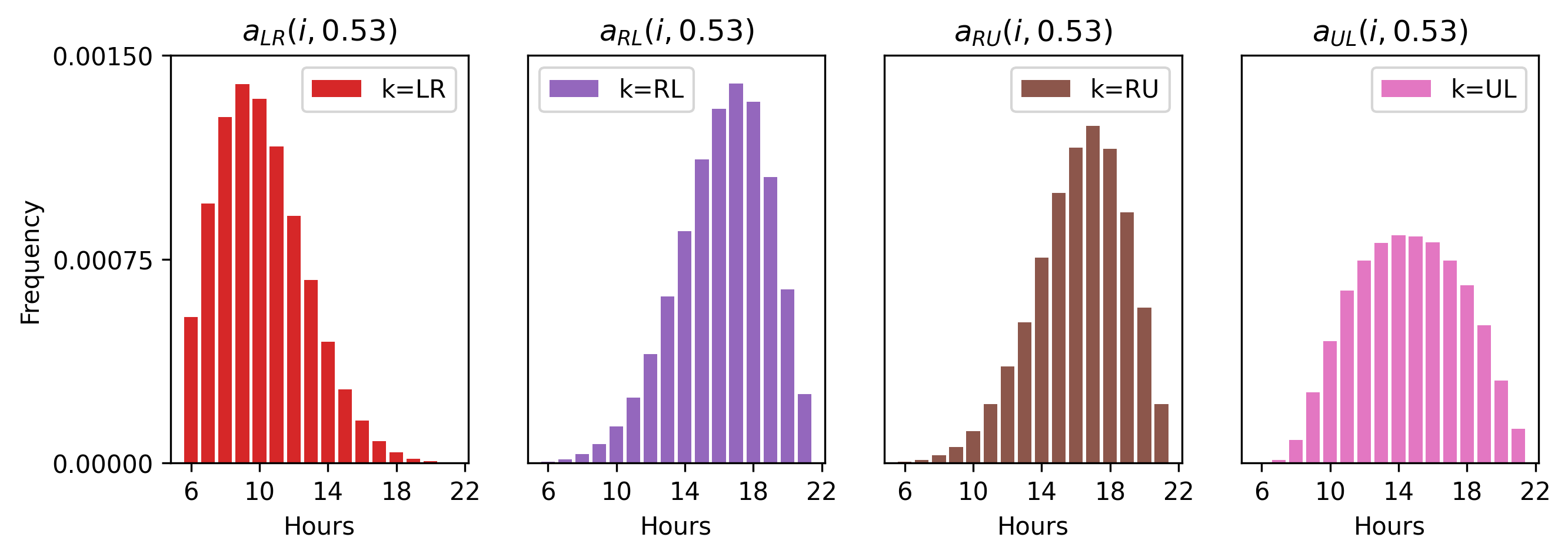}
    \caption{(Empirical approximation of the) attendance function at location $x=0.53$ for each journey type and each hour of the day, estimated from hidden data of Figure~\ref{fig ex hidden data}. For instance, the attendance function associated to $LR$ (red) tells us that individuals leaving $L$ for $R$ pass through $x=0.53$ more frequently in the morning than in the evening.}
    \label{fig ex att1}
\end{figure}

\begin{figure}[th]
    \includegraphics[width=1\textwidth]{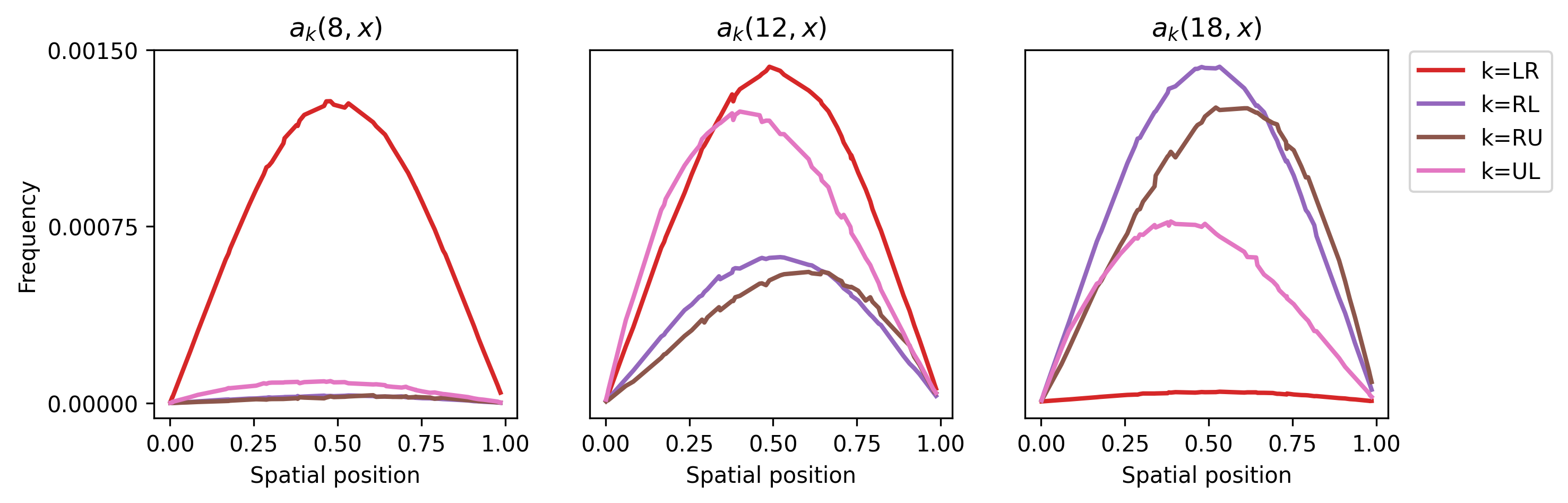}
    \caption{(Empirical approximation of the) attendance as a function of the counter location for each journey type and some hour of the day, estimated from hidden data of Figure~\ref{fig ex hidden data}. For instance, any counter will capture mainly $LR$ trips at 8 a.m., but only trips of the three other journey types at 6 p.m.}
    \label{fig ex att2}
\end{figure}

Given both vector of subpopulation sizes by journey and hidden data for 300 different days, we can define, in light of formula \eqref{empirical_attendance}, (an empirical approximation of) the attendance function $a_k(i,x_j)$, $k\in\{LR,RL,RU,UL\}$, $i\leq 24$, $j \leq 50$ as
\begin{equation}\label{attendance_function_guiding_example}
    a_k(i, x_j) : = \frac{1}{300}\sum_{d =1}^{300}\frac{ n_k^{(d)}(i, x_j)}{N_k^{(d)}}.
\end{equation}
For each activity, we can take a look at the values of $a_k$, either at a specific counter's location $x_j$ for each hour, as in Figure~\ref{fig ex att1}, or at a specific hour across each counter's location, as in Figure~\ref{fig ex att2}.

By construction, the hidden data and the model scaling parameters are known in this example. In what follows, we will simulate these hidden data for another day (which will not have been used to construct the attendance functions \eqref{attendance_function_guiding_example}) and will aim to estimate the scaling parameters from the aggregated counting data \eqref{aggregated_sum} together with the attendance functions previously estimated.

\section{Statistical model}\label{statistical_model}

To estimate the unknown vector of subpopulation sizes $N$, we propose a statistical model, associated with a likelihood function, which depends on the attendance functions $a_k(i, x)$, as well as the count data $n(i, x_j)$, and takes as input a vector of subpopulation sizes $\nu \in \RR_+^K$. The chosen estimator for $N$ will be the Maximum Likelihood Estimator (MLE).

The accuracy of the estimation should improve as the available counting data grow in size. It is not possible to improve the time resolution of the counters as the counting time intervals (indexed by $i$) are inherent to the devices. We thus choose to expand the dataset by increasing the number of counters, which can be implemented by a municipality. We present asymptotic convergence results (law of large numbers, central limit theorem) derived from the statistical model, and we are able to assess the estimation error and evaluate the effectiveness of the chosen spatial arrangement of the counters.

\subsection{Statistical framework} \label{statistical_hypothesis}
Given a journey type $k \leq K$, we model $n_k(i, x)$ (the number of passages in $x$ at time $i$) as a Poisson random variable of parameter (mean) $N_k a_k(i, x)$, introduced as the expected number of passages by $x\in\mathcal{E}$ at time step $i\leq I$,
\begin{equation}\label{law_hidden_data}
    n_k(i, x) \sim \mathcal{P}\left(N_k a_k(i, x)\right).
\end{equation}
Because of  the stability property of independent Poisson variables under summation,  the aggregated data $n(i, x)$ defined by \eqref{aggregated_sum} also follows a Poisson distribution,
\begin{equation*}
n(i,x) \sim \mathcal{P}(\langle N, a(i, x) \rangle).
\end{equation*}
From the perspective of equation \eqref{law_hidden_data}, Poisson noise can be seen as a random perturbation of the theoretical model given by $N_k a_k(i, x)$. From the aggregated formula, it can also be interpreted as measurement noise associated with the counter.

Poisson variables $n_k(i,x)$ are assumed to be independent from all others $n_{k'}(i', x')$. However, they are not identically distributed, as their parameter varies with the time interval of counting $i$ and with the location $x$. To achieve a more consistent statistical description of the data, we assume that $J$ counters are positioned randomly in $\left( x_j \right)_{j \leq J} \in \mathcal{E}^J$, where each $x_j$ is drawn independently according to some probability density $f_c$.

For each counter $j \leq J$, we define the variables $y_j$, which consist of both  position $x_j$ and counts $n(i, x_j)$,
\begin{equation}\label{random_variable_y_j}
    y_j := \left( x_j, \left(n\left(i, x_j\right) \right)_{i \leq I} \right).
\end{equation}
Since the locations $(x_j)_{j\leq J}$ are independent and identically distributed, the same holds for the sequence $(y_j)_{j \leq J}$. The randomization of counters position eliminates spatial dependence, while aggregating all time-dependent data per counter removes temporal dependence. The likelihood associated with $y_j$ is derived by conditioning on $x_j$,
\begin{equation}\label{y_j_likelihood}
    \nu \in \RR_+^K \mapsto \mathcal{L}(\nu, y_j) := f_c(x_j) \prod_{i = 1}^I \exp\Big(-\big \langle \nu, a(i, x_j) \big \rangle \Big) \,  \frac{\langle \nu, a(i, x_j) \rangle^{n(i, x_j)}}{n(i, x_j) ! },
\end{equation}
and, by independence, the likelihood of the dataset $\left( y_j \right)_{j \leq J}$ is given by
\begin{equation}\label{complete_likelihood}
    \nu \in \RR_+^K \mapsto \mathcal{L}(\nu, y_1, \dots, y_J) := \prod_{j = 1}^J \mathcal{L}(\nu, y_j).
\end{equation}
Given the explicit nature of the likelihood, the MLE naturally arises as a relevant method for estimating $N$. This approach is theoretically justified by asymptotic arguments in the following section, after which we propose an efficient algorithm for its computation.

\subsection{Asymptotic results in number of counters} \label{max_likelihood_subsection}

We now investigate the asymptotic properties of the MLE as the number of counters $J$ increases. Our goal is to verify that the MLE for this model is consistent and asymptotically Gaussian, mirroring classical statistical parametric frameworks \cite{van_der_vaart}. Only the results are presented here; detailed proofs are deferred to Appendix~\ref{appendix_stat}. 

From now on, we assume that for every $k \leq K$, the attendance function $a_k$ is strictly positive and continuous with respect to its spatial variable $x\in\mathcal{E}$. In addition, the collection of functions $(a_k)_{k \leq K}$ forms a linearly independent family.

The MLE for the observations $( y_j)_{j\leq J}$, denoted by $\widehat{N}_J \in\RR_+^K$, is defined as a maximizer of the likelihood function \eqref{complete_likelihood}. It should be noticed that $\widehat{N}_J = 0_K$ if and only if $n(i, x_j) = 0,$ for all $i \leq I$ and $j \leq J$. In this case, the likelihood reduces to a Dirac measure at $0_K$. Otherwise, the likelihood is strictly positive, smooth, and concave. We want to establish consistency and asymptotic normality of the MLE. We therefore assume that the true parameter $N\in \RR_+^K\setminus 0_K$. This guarantees that beyond a certain number of counters $J$, the MLE $\widehat{N}_J$ also lies in $\RR_+^K \setminus 0_K$ almost surely.

The concavity of the likelihood associated to the assumptions on the attendance functions and $N$ lead to the uniqueness and consistency of the MLE, as $J$ goes to infinity,
\begin{equation*}
    \widehat{N}_J \stackrel{\text{\tiny a.s.}}{\longrightarrow} N.
\end{equation*}

Before establishing asymptotic normality, we introduce the Fisher information matrix $\mathcal{I}(N)$, defined as the covariance matrix of the score function (gradient of the log-likelihood). In our model, it admits the following closed-form expression,
\begin{equation}\label{eq:def:fisher}
    \mathcal{I}(N):= \VV(\nabla_{\!N}\log \mathcal{L}(N, y) ) = \Biggl( \sum_{i=1}^I\int_{\mathcal{E}} \frac{a_k(i, x)a_{k'}(i, x)}{\langle N, a(i, x) \rangle}
                        f_c(x)dx 
                \Biggr)_{k, k' \leq K}.
\end{equation}
In particular, $\mathcal{I}(N)$ is invertible. The asymptotic normality takes the form of the following convergence, as $J$ goes to infinity,
\begin{equation*}
    \sqrt{J}\left(\widehat{N}_J-N\right) \stackrel{\text{\tiny d}}{\longrightarrow} 
    \mathcal{N}\left(0, \mathcal{I}(N)^{-1}\right).
\end{equation*}
This central limit theorem classically yields  confidence regions for $N$, as $J$ goes to infinity,
\begin{equation}\label{confidence_interval_ellipsoid}
    \PP\left(
        \left(\widehat{N}_J  -N\right)^\top 
            \mathcal{I}(N) 
        \left(\widehat{N}_J -N\right) 
        < \frac{\chi_{K, 1-\epsilon}^2}{J}\right) 
   \to1-\epsilon,
\end{equation}
where $\chi_{K,1-\epsilon}^2$ denotes the quantile of level $1-\epsilon$ of the Chi-square distribution with $K$ degrees of freedom. In other words, $N$ belongs to the inner ellipsoid centered in $\widehat{N}_J$ given by the positive definite matrix $J\,\mathcal{I}(N)/\chi_{K, 1-\epsilon}^2$ with probability $1-\varepsilon$ as $J$ goes to infinity.

\section{Approximation of the maximum likelihood estimator}\label{section_em}

We have established in the Section~\ref{statistical_model} that the MLE provides a consistent estimator of $N$, and its Fisher information accurately quantifies the estimation error, depending on the attendance function, but also on the distribution of counters. However, no closed-form expression is available for this estimator. For this reason, we will seek an approximation of the MLE, by adapting the Expectation-Maximisation (EM) algorithm to our framework. This will enable us to illustrate the results of Section~\ref{statistical_model} with our guiding example.

\subsection{Adapted version of the Expectation-Maximisation algorithm}

In Section~\ref{Counters_and_(des)agregation}, we introduced the hidden data $(n_k(i, x_j))_{i \leq I, j \leq J}$, which denote the number of passages associated to activity $k$ at each time interval $i$ and counter's location $x_j$. As a reminder, a counter positioned in $x_j$ can not observe directly these hidden data: instead, during the time interval $i$, the counter records their aggregated sum $n(i, x_j)$ given in \eqref{aggregated_sum}. As in equation \eqref{random_variable_y_j} where the dataset $(y_j)_{j \leq J}$ was introduced, we define now the hidden data $\left(y_j^\text{\tiny h}\right)_{j \leq J}$ where we replace the count data $n(i, x_j)$ by the hidden data $(n_k(i, x_j))_{k \leq K}$,
\begin{equation*}
    \left(y^\text{\tiny h}_j \right)_{j \leq J} := \left( x_j, \left( n_k(i, x_j) \right)_{k \leq K, i \leq I} \right)_{j \leq J}.
 \end{equation*}

We assume in Subsection~\ref{statistical_hypothesis} that the hidden counts follow Poisson distributions \eqref{law_hidden_data}. Therefore, this family is for the same reason as for $(y_j)_{j\leq J}$ an independent and identically distributed family of random variables, associated with an explicit likelihood function,
\begin{equation}\label{likelihood_hidden_data}
\nu \in \RR_+^K \mapsto \mathcal{L}^\text{\tiny h}(\nu, y^\text{\tiny h}_1, \dots, y^\text{\tiny h}_J) := 
\prod_{k=1}^K\prod_{j=1}^J \left\{f_c(x_j)\prod_{i=1}^I  
                \exp\left(-\nu_k a_k(i, x_j) \right) \frac{\left(\nu_k a_k(i, x_j)\right)^{n_k(i, x_j)}}{n_{k}(i, x_j)!}\right\}.
\end{equation}

The following ad hoc version of the EM algorithm \cite{dempster1977maximum} approximates the MLE by decomposing the aggregated sum \eqref{aggregated_sum} in iterative steps. It should be noted that $\EE_\nu^\text{\tiny h}$ (introduced in step 4) represents the conditional expectation of hidden variables under the parameter $\nu$, \textit{i.e.} with distribution $\mathcal{L}^\text{\tiny h}(\nu,\cdot)$.
\begin{algorithm}
\caption{Approximation of $\widehat{N}_J$}
\begin{algorithmic}[1]
\State $\forall\,i\leq I$, $\forall\,j\leq J$, generate $(z_{k,i,j}))_{k \leq K}$ randomly in $\NN^K$ so that $\sum_{k=1}^K z_{k,i,j} = n(i, x_j)$\hfill\texttt{initialization}
\State $\nu \gets \argmax_{\alpha \in \RR_+^K}  \mathcal{L}^\text{\tiny h}\left( \alpha, \{x_j, (z_{k,i,j}))_{k \leq K, i \leq I}\}_{j \leq J}\right)$ 
\While{$\nu$ evolves}
    \State $\{x_j,(z_{k,i,j})_{k \leq K, i \leq I} \}_{j \leq J} \gets \EE^\text{\tiny h}_\nu\left[ \left. \left( y^\text{\tiny h}_j \right)_{j \leq J} \right| \left(n(i, x_j)\right)_{i \leq I, j \leq J} \right]$\hfill\texttt{E step}

    \State $\nu \gets \argmax_{\alpha \in \RR_+^K}  \mathcal{L}^\text{\tiny h}\left( \alpha, \{x_j, (z_{k,i,j}))_{k \leq K, i \leq I}\}_{j \leq J}\right)$ \hfill\texttt{M step}
\EndWhile
\State return $\nu = (\nu_1, \dots, \nu_K)$
\end{algorithmic}
\label{algo_em}
\end{algorithm}

The strict concavity of the likelihood for a sufficient number of counters $J$ shown in Appendix~\ref{appendix_stat} guarantees the convergence of our EM algorithm to $\widehat{N}_J$, independently of the initial chosen data \cite{dempster1977maximum}. For implementation, explicit expressions for E and M steps are derived in Appendix~\ref{appendix_em} and are summed up below.
\begin{itemize}
    \item E step: the conditional expectation is given in closed-form by
    \begin{equation}\label{eq:formula:estep}
        \EE^\text{\tiny h}_\nu \left[ \left. \left(y_j^\text{\tiny h} \right)_{j \leq J}
            \right| 
            \left(n(i, x_j) \right)_{i \leq I, j \leq J} \right]
        = \left( x_j, \left(n(i, x_j) \frac{\nu_k a_k(i, x_j)}{\langle\nu, a(i, x_j) \rangle} \right)_{k \leq K, i \leq I} \right)_{j\leq J}.
    \end{equation}
    \item M step: the maximum of the (hidden) likelihood is given in closed-form by
    \begin{equation}\label{eq:formula:mstep}
    \argmax_{\nu \in\RR_+^K}  
        \mathcal{L}^\text{\tiny h} \left( \nu, 
            \left( y_j^\text{\tiny h} \right)_{j \leq J}\right) 
    = \left( \frac
        {\sum_{i\leq I, j\leq J} n_k(i, x_j)}
        {\sum_{i\leq I, j\leq J} a_k(i, x_j)} 
        \right)_{k \leq K}.
    \end{equation}
\end{itemize}

\subsection{Guiding example: estimation results}\label{ss:example:results}

\begin{figure}
    \centering
    \includegraphics[width=1\textwidth]{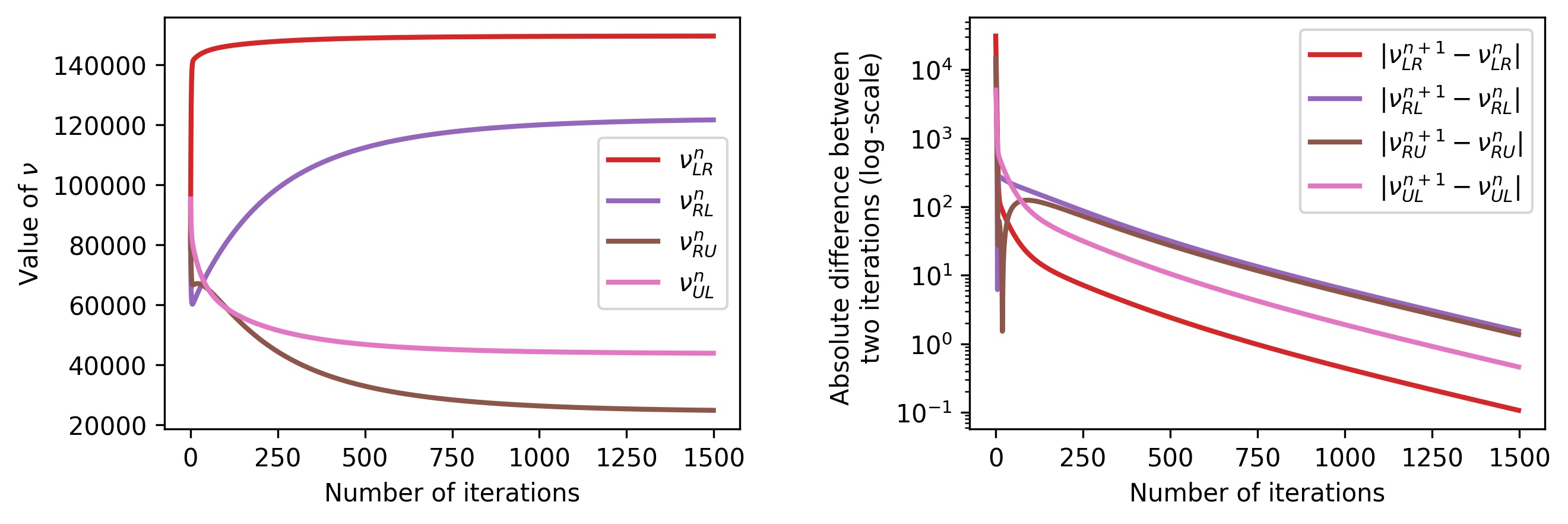}
    \caption{Convergence of our EM algorithm toward the MLE (left) and absolute difference in log-scale between two iterations of the algorithm (right), showing an exponential convergence.}
    \label{fig:convergence:em}
\end{figure}

We aim to illustrate our theoretical and algorithmic statistical results in the framework of the guiding example introduced in Subsection~\ref{guiding_example_intro}. To this end, we generate counting data on a new day, with a known vector of subpopulation sizes $N=(N_{LR}, N_{RL}, N_{RU}, N_{UL}) = (150\,000, 120\,000, 30\,000, 40\,000)$. By using only the new counting data $n(i, x_j)_{i \leq 24, j \leq 50}$ and the attendance functions defined from prior data by formula \eqref{attendance_function_guiding_example} (and presented in Figures~\ref{fig ex att1} and \ref{fig ex att2}), our main objective is to estimate the scaling parameter $N$. To this end, we run our EM algorithm designed to approximate the MLE. Figure~\ref{fig:convergence:em} illustrates the exponentially-fast convergence of our optimization method. 

\begin{figure}
    \centering
    \includegraphics[width=.8\textwidth]{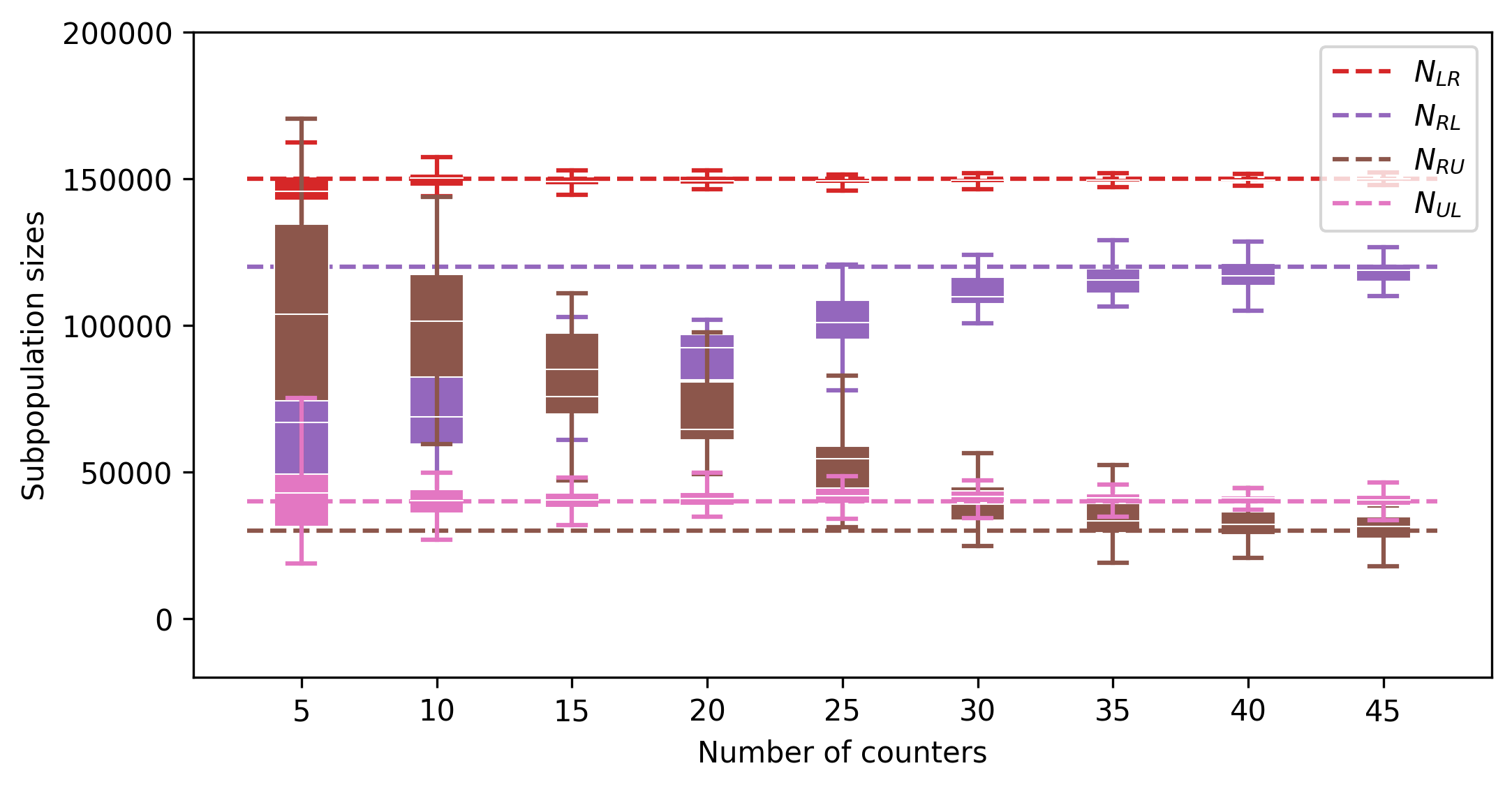}
    \caption{Boxplots of the MLE for the four components of $N$ (in different colors) over fifty independent replicates (days) for different numbers of counters.}
    \label{consistency_toy_example}
\end{figure}

\begin{figure}
    \centering
    \includegraphics[width=1\textwidth]{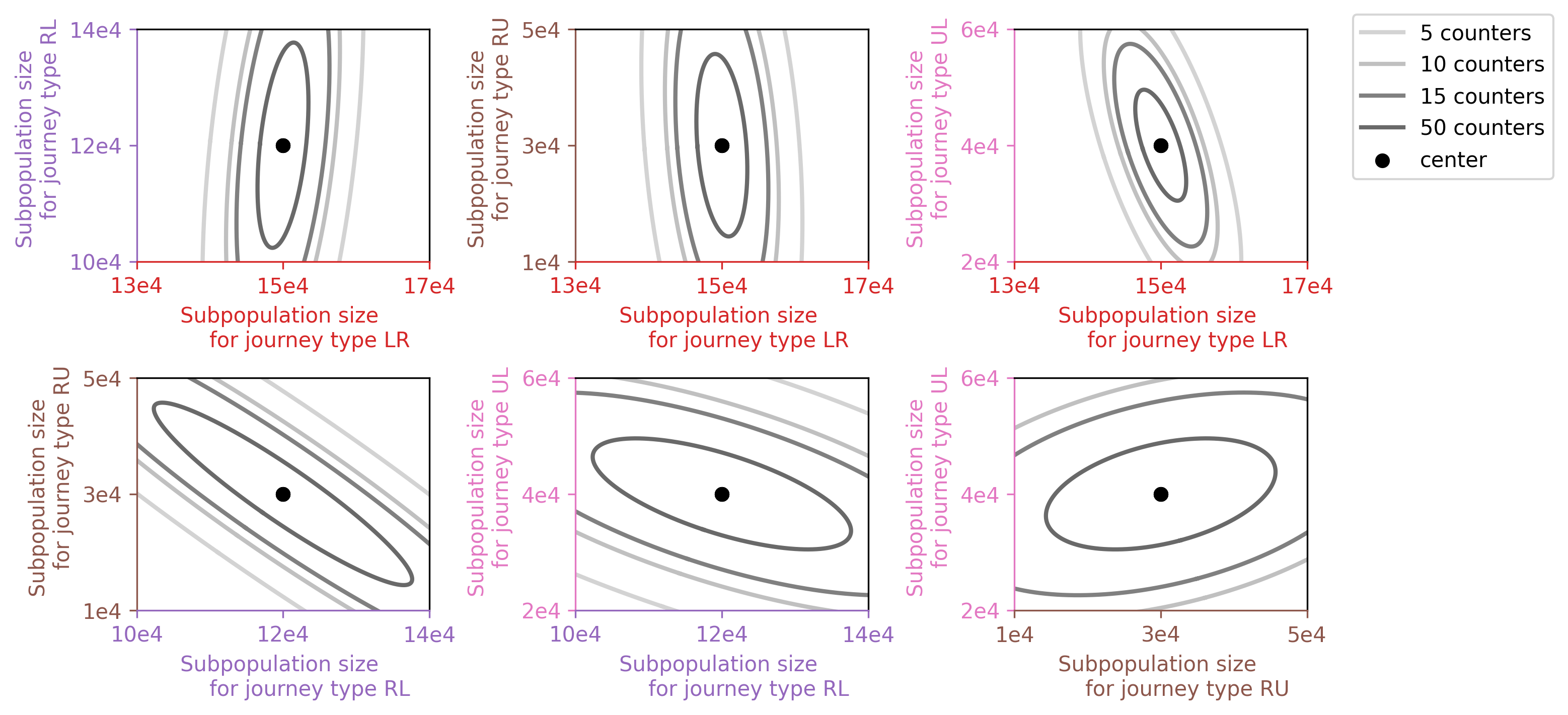}
    \caption{The six different 2D slices of the 4D confidence ellipsoids at $95\%$ confidence level, corresponding to $5$, $10$, $15$, and $50$ counters. The lengths of the ellipsoid axes decrease inversely with the square root of the number of counters.}
    \label{figure confidence ellipsoid}
\end{figure}

We now illustrate the theoretical results of Section~\ref{statistical_model} on the MLE. For the same vector of subpopulation sizes $N$ as in the previous paragraph, we generate trips across fifty days. For each day, we estimate the associated MLEs successively with an increasing number of counters, from five up to fifty. The distribution of the MLEs is given as boxplots in Figure~\ref{consistency_toy_example}, illustrating the convergence of $\widehat{N}_J$ toward $N$ when $J$ increases. With $J=30$ counters, all components of $N$ are well estimated. The asymptotic confidence regions given in \eqref{confidence_interval_ellipsoid} take the form of a 4D ellipsoids in this illustrative example. We present 2D slices of these confidence regions at the $95\%$ confidence level and from up to fifty counters in Figure~\ref{figure confidence ellipsoid}.

It should be noted that the shape and the size of the confidence regions \eqref{confidence_interval_ellipsoid} depend on the distribution of counters. For a fixed number of counters, it is relevant to locate them such that the corresponding confidence regions are as small as possible, to improve the accuracy of the overall method. Here, we try out different strategies for positioning the counters. Up to this point, in this guiding example, the counters were distributed uniformly. We now add a second strategy, by placing counters primarily along the boundaries, where one of the activities ($L$ or $R$) predominates over the other. We also evaluate a third strategy, with counters concentrated on the center of the area, where highest diversity of journey types occurs simultaneously. 

\begin{figure}
    \centering
    \includegraphics[width=.9\textwidth]{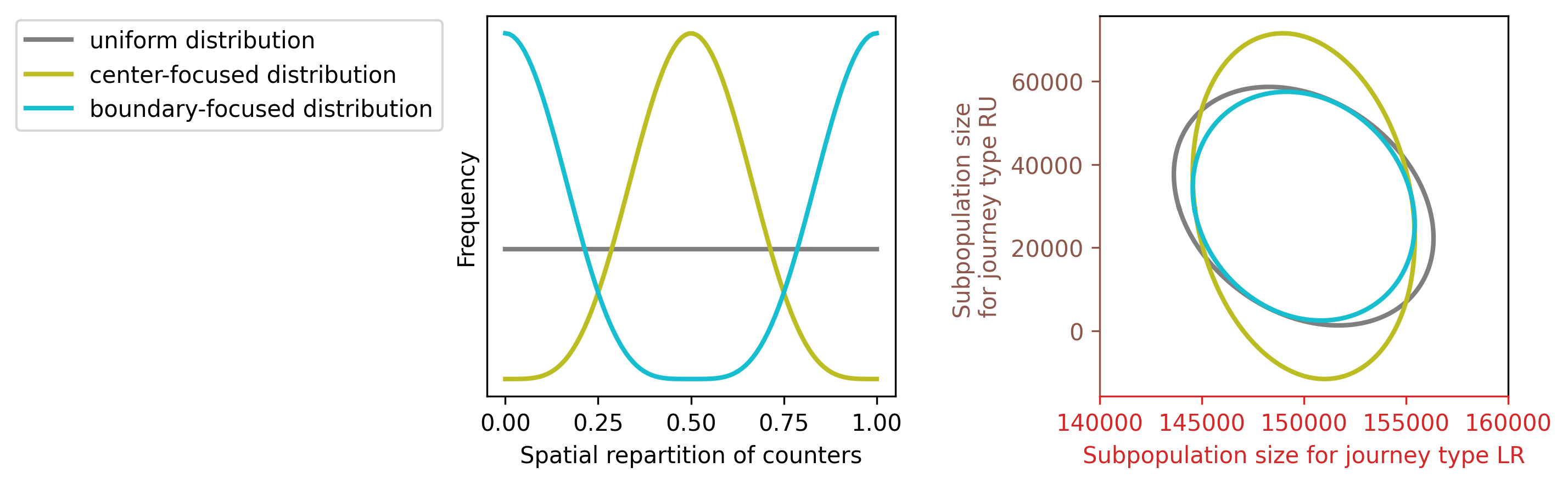}
    \caption{Three distribution strategies (in different colors) for positioning the counters over the spatial domain $[0,1]$ (left) and one of the six 2D slices (for journey types $RU$ and $LR$) of the 4D $95\%$ confidence ellipsoids estimated from fifteen counters, corresponding to each strategy (in different colors).}
    \label{fig strategies counters}
\end{figure}

More precisely, the three selected distribution over $[0, 1]$ are $d_u : x \mapsto 1$, $d_b : x \mapsto 8 \cos(\pi x)^4/3$, and $d_c : x \mapsto 8\sin(\pi x)^4/3$. Alongside the graph of these distribution, we present in Figure~\ref{fig strategies counters} an example of 2D slices of the associated confidence ellipsoids for fifteen counters, for the three counter distribution strategies. These simulation results show that, in our example, the best strategy is to place counters where one of the activity is highly represented. Up to our knowledge, it is difficult to have an intuition on which strategy yields better results. Our statistical framework provides a strategy to deduce relevant answer, and appears to be a promising avenue for developing cost-effective solutions of mobility dynamics analysis. 

%%%%%%%%%%%%%
% S 5
%%%%%%%%%%%%%
\section{Modeling the attendance function from individual journeys}  \label{stochastic_deterministic}

In this section we design, in a one-dimensional setting, the attendance function for a given journey type using geographic data related to both activities involved (spatial distribution of origin and destination activities and their temporal schedules). We begin with a stochastic, individual-based model of trips between activities. For a large population, we capture a spatio-temporal flux function. When evaluated at a point $x$ and integrated over the $i$-th time step, it yields the attendance function $a(i, x)$. This flux function can be interpreted as the sum of oriented fluxes, which are solutions to classical transport equations with source and death terms, analogous to some ecological models. The transport term describes the movement of individuals, while the source term accounts for individuals leaving their previous activity, \textit{i.e.} entering the study area $\mathcal{E}$. The death term, on the other hand, represent individuals arriving at their next activity, \textit{i.e.} exiting the study area $\mathcal{E}$. Two variants of the model are considered: the first, where schedules are governed by departure times, is described in Subsections~\ref{single_individual} and \ref{large_population}; the second, where schedules are defined by arrival times, is addressed in Subsection~\ref{alternative_model}. The mathematical details for the latter model are relegated to Appendix~\ref{appendix_attendance}.

%%%%%%%%%%%%%
% SS 5.1
%%%%%%%%%%%%%
\subsection{Individual based model}\label{single_individual}

\begin{figure}[ht]
	\centering	
	\def\svgwidth{.8\textwidth}
	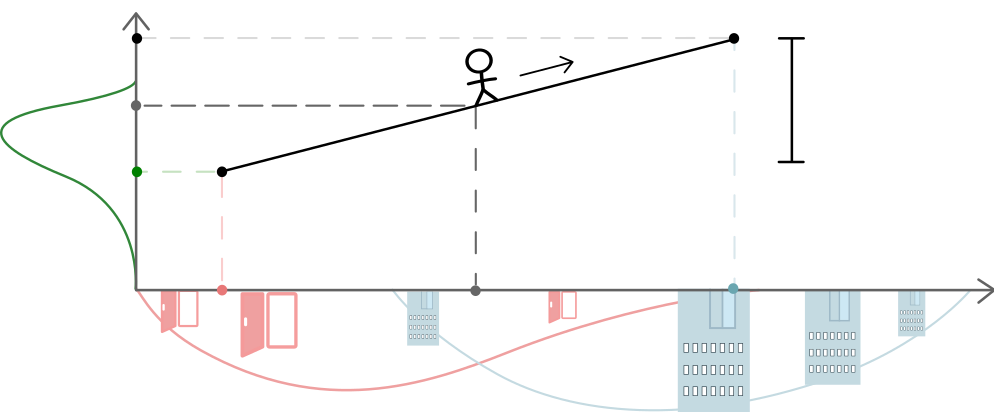
	\vspace{-.3cm}
	\caption{Random trajectory of an individual $p$. It begins at $x_p^0$, drawn from the red distribution (doors) and arrives at $x_p^e$, drawn from the blue distribution (buildings). The green distribution, used to draw the starting time $t_p^0$, depends on $x_p^0$. The velocity $v_p$ is also drawn from a distribution $f_v$ (not shown in the graphic). At time $t \in \mathcal{T}_p$, the individual is located at $\gamma_p(t)$.}
	\label{fig:diagram:trajectory}
\end{figure}

For a given journey type $k$, consider $N_k$ individual trajectories over the one dimensional spatial domain $\mathcal{E} = [x_l, x_r]$, and during the time interval $T \subset \RR$. This interval is further divided in $I$ disjoint time steps $\left(T_i\right)_{i\leq I}$, such that  $T = \bigcup_{i \leq I} T_i$. Each individual trajectory of type $k$, indexed by $p \leq N_k$, describes the continuous, straight-line path of an individual from its origin to its destination. It consists in its position as a function of time. Each trajectory is drawn according to the following random variables whose distribution only depend on the journey type $k$:
\begin{itemize}
	\item $v_p$, the velocity, with distribution $f_v$. Each individual travels at a constant velocity, that may vary across individuals. For pedestrians, explicit distributions can be found for example in \cite{chandra2013speed}.  
	\item $x^0_p$, the starting point, with distribution $f_{x^0}$. It is the location of the previous activity of the individual, \textit{i.e.} before moving. For instance, for work-to-home journeys, $f_{x^0}$ may be modeled as the density of offices.  The cumulative distribution function and survival function of $x^0_p$ are denoted by $F_{x^0}$ and $S_{x^0}$. As a reminder, by definition $S_{x^0} := 1-F_{x^0}$.
	\item $t^0_p$, the starting time, \textit{i.e.} the time at which the individual leaves its previous activity. We further assume that this starting time may also depend on the starting point, $x^0_p$. The conditional distribution of starting time has a density (with respect to the Lebesgue measure), $t  \mapsto f_{t^0|x^0}(t | x)$, for almost every $x \in \mathcal{E}$. For work-to-home commutes, this conditional density represents the end-of-work schedules at each location. The joint distribution of $(t^0_p, x^0_p)$ is written as $f_{(t^0, x^0)}$. 
    \item $x^e_p \in \mathcal{E}$, the destination point, with distribution $f_{x^e}$. It is the location of the next activity of the individual. Its distribution function is denoted by $F_{x^e}$ and its survival function by $S_{x^e}$. For work-to-home commutes, this represents the density of homes\footnote{\url{https://datafoncier.cerema.fr/sites/datafoncier/files/fichiers/2019/02/CETE\_NP\_carte\_densite\_logt\_carroyage\_100m\_A1pdf\_0.pdf}, last consulted on April 12, 2026}.
\end{itemize}	
We assume that $v_p$, $(x^0_p, t^0_p)$ and $x^0_p$ are independent random variables. The individual $p$ stands at $x^0_p$ at time $t^0_p$ and moves towards $x^e_p$ with velocity $v_p$. Therefore, at time $t \geq t^0_p$, if it is still moving at this time, its  position is given by
\begin{equation}\label{def traj}
    \gamma_p(t) := x^0_p + \epsilon_p (t-t_p^0)v_p,
\end{equation}
with $\epsilon_p = 1$ if $x^e_p \geq x^0_p$ and $\epsilon_p=-1$ else. More precisely, it reaches its destination (to start its new activity) at time $\gamma_p^{-1}(x^e_p) = t^0_p + |x^e_p - x^0_p|/v_p$. The trajectory of the individual $p$ is thus defined as the restriction of $\gamma_p$ over the time interval 
\begin{equation*}
	\mathcal{T}_p := \left[ t^0_p, t^0_p + \frac{|x^e_p - x^0_p|}{v_p} \right].
\end{equation*}
The reader is referred to Figure~\ref{fig:diagram:trajectory} for an explanatory diagram.

%%%%%%%%%%%%%
% SS 5.2
%%%%%%%%%%%%%
\subsection{Large population model}\label{large_population}
We still consider a fixed journey type $k$. Here we show that if the number of (random) individual trips is large, that is $N_k$ goes to infinity, then, at any given position $x$ and  during any counting interval $T_i$, $i \leq I$, the number of passages is deterministic. In other words, with enough trips, we can predict the exact number of people that a counter should be recording. More precisely, we show that this number can be written as the product $N_k \int_{t_{i-1}}^{t_i} \nu_k(t, x)dx$. We identify $\nu_k$ as a non-oriented flux, that depends only on geographic data $f_v$, $f_{(t^0, x^0)}$ and $f_{x^e}$.

Assume that all individuals behave independently from each other. For example, for pedestrians, this is true with a low density of individuals,  or anytime that an individual's trajectory is not hindered by surrounding pedestrians. In that case, the family of random variables $\left( \gamma_p, \mathcal{T}_p \right)_{p\leq N_k}$ is independent and identically distributed. We consider $n_k(i, x)$ the number of individuals passing through position $x \in \mathcal{E}$ during the time interval $T_i$,
$$
	n_k(i, x) := \sum_{p=1}^{N_k}\ind_{\left\{x \in \gamma(T_i \cap \mathcal{T}_p)\right\}}.
$$	
Define $a_k(i, x) := \PP\left(x \in \gamma(T_i \cap \mathcal{T}_p) \right)$ the probability that a single individual $p$, with journey type $k$ passes through $x$ during $T_i$. The fact that the $\mathcal{T}_p$ are independent and identically distributed implies that $a_k(i, x)$ does not depend on $p$. The quantity $n_k(i, x)$ is a binomial variable with parameter $(N_k, a_k(i, x))$. The law of large  numbers thus guarantees the almost sure convergence,
\begin{equation*}
	\frac{n_k(i,x)}{N_k} \stackrel{\text{\tiny a.s.}}{\longrightarrow} a_k(i, x).
\end{equation*}

In Appendix~\ref{subsec app att1}, we show rigorously how probability $a_k(i, x)$ can be expressed as a the integral of an non-oriented flux function $\nu_k(t,x)$,
\begin{equation}\label{eq ak flux}
	a_k(i, x) = \int_{T_i} \nu_k (t, x)dt.
\end{equation}
where non-oriented flux $\nu_k(t, x)$ has the following explicit formula, for $t \geq 0$ and $x \in \RR$,
\begin{equation}\label{unoriented_flux}
	\nu_k(t, x):=  
        \int_{\RR_+^*} \int_\RR G_{x}(u)
	        f_{t^0|x^0} \left( \left. t - \frac{|u-x|}{v} \right| u \right)
	        f_{x^0}(du) f_v(dv).	
\end{equation}
It depends solely on the laws $f_v, f_{x^e}$ and $f_{(t^0, x^0)}$, through their survival and distribution function (denoted by $S$ and $F$, as in the previous section). The term $|u-x|/v$ represents the remaining travel time from $x$ to $u$ at speed $v$. Besides, the intermediary function $G_{x} $ captures the directionality of travel with respect to $x$, defined, for any $u\in\RR$, as the sum of two terms,
\begin{equation}\label{eq:def:Gx}
    G_{x}(u) := S_{x^e}(x) \ind_{]-\infty, x[}(u) + F_{x^e}(x) \ind_{[x, +\infty[}(u).
\end{equation}
Taking into account that $F_{x^e} +S_{x^e}=1$, each term in this definition contributes to one direction of travel. As a consequence, the flux function itself has two oriented components: a rightward flux $\nu_\rightarrow(t, x)$ and a leftward flux $\nu_{\leftarrow}(t, x)$, both of which depend solely on the laws $f_v, f_{x^e}$ and $f_{(t^0, x^0)}$,
\begin{align}
\nu_\rightarrow(t, x) &  = \int_{\RR_+^*} \left\{S_{x^e}(x)\int_{-\infty}^{x} 
                f_{t^0|x^0} \left( \left. t - \frac{x-u}{v} \right| u \right) 
                f_{x^0}(du) \right\} f_v(dv), \label{nu_right} \\
    \nu_\leftarrow(t, x) & = \int_{\RR_+^*} \left\{F_{x^e}(x) \int_{x}^{+\infty} 
                f_{t^0|x^0} \left( \left. t - \frac{u-x}{v} \right| u \right) 
                f_{x^0}(du) \right\} f_v(dv). \label{nu_left}
\end{align}
In the rightward flux, we consider the starting times $t - (x-u)/v$ for an individual passing through $x$ at time $t$, for any starting position $u$ located to the left of $x$. Similarly, in the leftward flux, we consider the starting time $t - (u-x)/v$ for an individual passing through $x$ at time $t$, for any starting position $u$ located to the right of $x$. The technical details are presented in Appendix~\ref{subsec app att1} for the alternative modeling approach, described in Subsection~\ref{alternative_model}, where schedules are governed by arrival times.

%%%%%%%%%%%%%
% SS 5.3
%%%%%%%%%%%%%
\subsection{Alternative model with specified arrival time} \label{alternative_model}

In the previous individual-based model above, we assumed total knowledge of the distribution of starting times, for a given journey type. However, depending on the activity, geographic data sometimes relates to arrival times rather than starting time, in particular for home-to-work trips. Therefore, if work schedules are used to design the model (see for example \cite{chenu2002horaires,colomb2025generation}), it becomes necessary to derive the equations differently. Our approach still works seamlessly in this case, leading once again to an explicit formula similar to \eqref{unoriented_flux}.

We shift slightly the modeling paradigm that lead to \eqref{def traj}. First, instead of the starting time, we consider the arrival time depending on the arrival position: the arrival time of an individual $p$ conditionally to $x^e_p = x$ is assumed to admit a density (with respect to the Lebesgue measure) $t\mapsto f_{t^e|x^e}(t |x)$, for almost every $x \in \mathcal{E}$. The joint distribution of $(t^e_p, x^e_p)$ is written as $f_{(t^e, x^e)}$. 

This leads to define the analogous trajectory to \eqref{def traj}, specific to our new assumptions: for travel time $t$, individual $p$ locates at position 
\begin{equation}\label{gamma_arrival_time}
    \gamma_p(t) = x^e_p - \epsilon_p (t^e_p-t)  v_p, 
\end{equation}
with $\epsilon_p = 1$ if $x^e_p \geq x^0_p$ and $\epsilon_p=-1$ else. The time interval of the journey is precisely
\begin{equation*}
	\mathcal{T}_p := [\gamma_p^{-1}(x^0), t^e] = \left[t^e_p - \frac{|x^e_p - x^0_p|}{v_p}, t^e_p \right].
\end{equation*}
With those notations, for any fixed journey type $k$, $n_k(i,x)/N_k$ almost surely goes to $a_k (i, x)$, and this attendance function is the sum of two directional fluxes,
\begin{equation}\label{eq attendance arrival time}
	a_k(i, x)   = \int_{T_i} \nu_\rightarrow(t, x) dt + \int_{T_i} \nu_\leftarrow(t, x) dt.
\end{equation}
The difference lies in the definition of the rightward flux $\nu_\rightarrow$ and of the leftward flux $\nu_\leftarrow$. We show in Appendix~\ref{subsec app att1} that they are defined as
\begin{align}
     \nu_\rightarrow(t, x) &:= 
            \int_{\RR_+^*} \left\{ F_{x^0}(x)\int_x^{+\infty} 
                f_{t^e|x^e} \left( \left. t + \frac{u-x}{v} \right| u \right) 
                f_{x^e}(du) \right\}f_v(dv), \label{nu_right_alternative}\\
    \nu_\leftarrow(t, x) &:= 
            \int_{\RR_+^*} \left\{S_{x^0}(x) \int_{-\infty}^x 
                f_{t^e|x^e} \left( \left. t + \frac{x-u}{v} \right| u \right)
                f_{x^e} (du) \right\}f_v (dv) \label{nu_left_alternative}. 
\end{align}
The differences between these definitions of oriented fluxes and their analogous with starting time distribution \eqref{nu_right} and \eqref{nu_left} stem to trajectory considerations: for example, to obtain the rightward flux \eqref{nu_right_alternative}, we consider the arrival time $t + (u-x)/v$ of an individual passing through $x$ at time $t$, for any arrival position $u$ located to the right of $x$. In contrary, in the previous model \eqref{nu_right}, we considered the starting time $t - (x-u)/v$ for an individual passing through $x$ at time $t$, for any starting position $u$ to the left of $x$.

%%%%%%%%%%%%%
% SS 5.4 Guiding example
%%%%%%%%%%%%%
\subsection{Guiding example: derivation of attendance functions} \label{ss:example:theory}

The guiding example introduced in Subsection~\ref{guiding_example_intro} is defined from counting data where the journey type is observed while it is usually hidden (see Figure~\ref{fig ex hidden data}). We explain here how these hidden data were actually generated. For each journey type $k \in \{LR, RL, RU, UL \}$, a theoretical attendance function $a_k(i,x)$ is defined using formulas \eqref{unoriented_flux}, \eqref{nu_right} and \eqref{nu_left}, from the spatial distribution of activities and temporal schedules given in Figure~\ref{spatial_distribution_agenda}, assuming a constant velocity $v=2$ for each individual. In other words, the distributions of $x^0$, $t^0$ and $x^e$ are as depicted in Figure~\ref{spatial_distribution_agenda}, and a Dirac velocity distribution $f_V = \delta_{2}$ is selected in \eqref{nu_right} and \eqref{nu_left}. For example, for journey type $k=LR$, $f_{x^0}$ is the spatial distribution of activity $L$ (represented by the dotted curve in Figure~\ref{spatial_distribution_agenda} (left)), $f_{x^e}$ is the spatial distribution of activity $R$ (represented by the dashed curve in Figure~\ref{spatial_distribution_agenda} (left)) and $f_{t^0}$ is the end time of activity $L$, and thus the starting time of the journey of type $LR$ (represented by the dotted curve in Figure~\ref{spatial_distribution_agenda} (right)). 

Then, at day $d$, given the vector of subpopulation sizes $N^{(d)}$ of Figure~\ref{fig ex nbtrips}, and using approximation \eqref{eq hidden data attendance}, we expect $N_k^{(d)} a_k(i, x_j)$ passages for journey type $k$ at time step $i$ and counter location $x_j$. To generate the synthetic hidden data $n_k^{(d)}(i, x_j)$ presented in Figure~\ref{fig ex hidden data}, we added an independent Poisson noise to each of these expected passages as in \eqref{law_hidden_data}, following the statistical framework described in Subsection~\ref{statistical_hypothesis}. Hence, the attendance functions presented in Figures~\ref{fig ex att1} and \ref{fig ex att2} are noisy versions of the theoretical attendances from formula \eqref{unoriented_flux}. As a consequence, the numerical experiments presented in Subsection~ \ref{ss:example:results} are done from data generated according to the statistical model assumed.

\begin{figure}[th]
    \centering
    \includegraphics[width=1\textwidth]{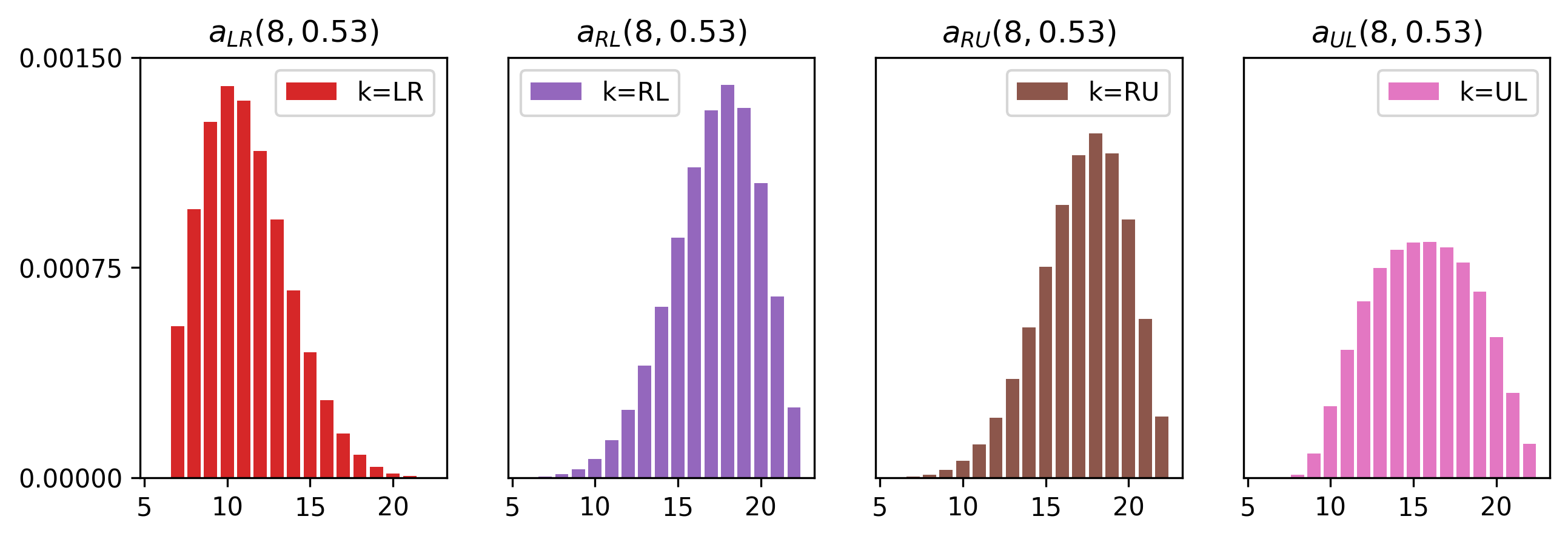}
    \caption{(Theoretical) attendance at location $x=0.53$ for each journey type and each hour of the day, computed from formulas \eqref{eq attendance arrival time}, \eqref{nu_right} and \eqref{nu_left} and spatial and temporal distributions of Figure~\ref{spatial_distribution_agenda}. The empirical attendance of Figure~\ref{fig ex att1} is a noisy version of this theoretical attendance presented in the same setting.}
    \label{fig theoretical attendance fixed location}
\end{figure}

\begin{figure}[th]
    \centering
    \includegraphics[width=1\textwidth]{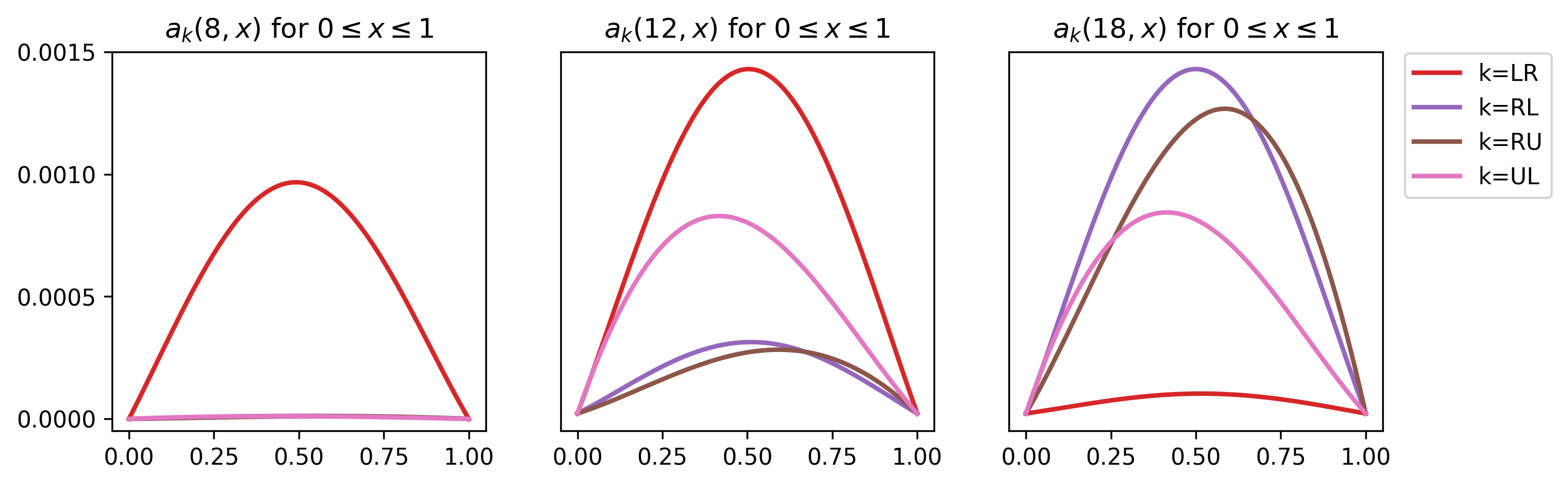}
    \caption{(Theoretical) attendance as a function of the counter location for each journey type and some hour of the day, computed from formulas \eqref{eq attendance arrival time}, \eqref{nu_right} and \eqref{nu_left} and spatial and temporal distributions of Figure~\ref{spatial_distribution_agenda}. The empirical attendance of Figure~\ref{fig ex att2} is a noisy version of this theoretical attendance presented in the same setting.}
    \label{fig theoretical attendance fixed times}
\end{figure}

Theoretical attendance functions $a_k(i, x)$ -- obtained from formulas \eqref{unoriented_flux}, \eqref{nu_right} and \eqref{nu_left} -- are presented in Figures~\ref{fig theoretical attendance fixed location} and \ref{fig theoretical attendance fixed times} in the exact same setting of their empirical versions of Figures~\ref{fig ex att1} and \ref{fig ex att2}.

%%%%%%%%%%%%%
% SS 5.5
%%%%%%%%%%%%%
\subsection{Interpretation of the flux: insight from mathematical biology}

If every individual walks at the same speed, we can interpret the flux as the (weak) solution to a transport equation with source and death terms inspired by mathematical ecology, as in \cite{murray2007mathematical} and references therein. In other words, we provide an interpretation of the bracketed terms in the oriented flux expressions \eqref{nu_right} and \eqref{nu_left} (in the starting time setting) and \eqref{nu_right_alternative} and \eqref{nu_left_alternative} (in the arrival time setting), which means at a given speed, before integration over the velocity distribution. In that spirit, if the distribution of speed is a Dirac mass, $\delta_v$, the flux of the model based on starting times of Subsections~\ref{single_individual} and \ref{large_population} can be written as 
\begin{align}
    \nu_\rightarrow(t,x) &= S_{x^e}(x) \int_{-\infty}^{x} f_{t^0 | x^0}\left( \left. t- \frac{x-u}{v} \right| u \right) f_{x^0}(du),\label{nu_droit_simplifie_ti} \\
    \nu_\leftarrow(t, x) &= F_{x^e}(x) \int_{x}^{+\infty} f_{t^0|x^0} \left( \left. t - \frac{u-x}{v} \right| u \right) f_{x^0}(du).\label{nu_gauche_simplifie_ti} 
\end{align}
For the model based on arrival times of Subsection~\ref{alternative_model}, one finds
\begin{align}
    \nu_\rightarrow(t,x) &= F_{x^0}(x) \int_x^{+\infty} f_{t^e | x^e}\left( \left. t+ \frac{u-x}{v} \right| u \right) f_{x^e}(du),\label{nu_droit_simplifie_tf} \\
    \nu_\leftarrow(t, x) &= S_{x^0}(x) \int_{-\infty}^x f_{t^e|x^e} \left( \left. t + \frac{x-u}{v} \right| u \right) f_{x^e}(du).\label{nu_gauche_simplifie_tf}
\end{align}
We examine the interpretation of equation \eqref{nu_droit_simplifie_ti} in more detail. For the other expressions, the same approach is of course also valid. First, recall that $S_{x^e}(x)$ represents the survival function of the random variable $f_{x^e}$, which stands for the time of arrival of the individual. By definition, this means that $S_{x^e}(x)$ is the probability that an individual moving to the right will travel to a destination located beyond $x$, \textit{i.e.} to the right of $x$. 

In Appendix~\ref{subsec app att2}, we show that the flux $\nu_\rightarrow$ defined in \eqref{nu_droit_simplifie_tf} is the solution of a transport equation with source and death terms ($b$ and $d$)  of the form
\begin{equation*}
    \partial_t \nu_\rightarrow(t, x) + v \partial_x \nu_\rightarrow(t, x) 
    = v b(t, x) - v d(t, x).
\end{equation*}
This is also the case of the flux $\nu_\rightarrow$ defined in \eqref{nu_droit_simplifie_ti}. We recognize a transport equation at speed $v$, which is consistent with a Dirac speed distribution. Furthermore, we have precise expressions for $b$ and $d$. We provide a detailed interpretation of these formulas in the context of the rightward flux associated to equation \eqref{nu_droit_simplifie_ti}.
The source term is  
\begin{equation*}
    b(t, x) =   S_{x^e}(x) f_{(t^0, x^0)}(t, x).
\end{equation*}
This means that new individuals begin a trip rightward, at rate $b(t,x)$. In order to do that, they must follow two conditions. First, they must be at the starting point of the activity at the corresponding time, which happens with probability   $f_{(t^0, x^0)}$. In addition, they must begin to travel towards a destination at their right. This is the reason for the appearance of the survival function $S_{x^e}(x)$ here. The individuals starting from point $x$ are partitioned into two groups: those who move left (with proportion $F_{x^e}(x)$) and those who move right (with proportion $S_{x^e}(x)$). This second group is the  one that starts their rightward trip, and therefore, enters the count. The other group is accounted for  in $\nu_\leftarrow$. 

On the other hand, the death term is written as
\begin{equation*}\label{eq death term}
    d(t, x) = \nu_\rightarrow (t, x)\frac{f_{x^e}(x)}{S_{x^e}(x)}.
\end{equation*}
Observe the appearance of the risk function $f_{x^e}(x)/S_{x^e}(x)$ (or more generally a risk measure, in the case where $f_{x^e}$  is only a Borel measure, without density with respect to the the Lebesgue measure),
\begin{equation*}
    \frac{f_{x^e}(x)}{S_{x^e}(x)} := \lim_{h \to 0}\frac{\PP(x^e < x+h|x^e\geq x)}{h}. 
\end{equation*}
It quantifies the instantaneous risk that the current location $x$ is the arrival location of the trajectory: those individuals are no longer traveling, and therefore exit the count.  The death term is also proportional to the flux $\nu_\rightarrow$: the risk function is the rate at which individuals disappear from the current traveling population.

We have therefore established that $\nu_\rightarrow$ verifies, in a weak sense, the following transport equation,
\begin{equation}\label{eqint edp flux}
    \partial_t \nu_\rightarrow(t, x)  + v\partial_x \nu_\rightarrow(t, x)
			= v S_{x^e}(x) f_{(t^0, x^0)}(t, x) -v \nu_\rightarrow(t, x)\frac{f_{x^e}(x)}{S_{x^e}(x)}.
\end{equation}
It should be noticed that we slightly abused notation in the right-hand side since  $f_{(t^0,x^0)}$, and  $f_{x^e}$ are not functions defined for all $(t,x)$ but probability measures. This explains why this equation is defined on the dual of $\mathscr{C}^\infty_c(\RR \times \RR)$, meaning that, for all $\psi \in \mathscr{C}^{\infty}_c(\RR \times \RR)$,
\begin{multline*}
  -   \int_{\RR^2}    \nu_\rightarrow (t, x) \Big[ \partial_t \psi(t, x) + v \partial_x \psi(t,x) \Big] dt\,dx \\ 
    =  \int_{\RR^2}  \psi(t, x) v S_{x^e}(x) f_{(t^0, x^0)}(dt, dx) 
    - \int_\RR \int_\RR \psi (t, x) v \nu_\rightarrow (t, x) \frac{1}{S_{x^e}(x)}f_{x^e}(dx) dt.
\end{multline*}
This expression is perfectly defined in our functional setting. Note that the  left-hand side is the classical weak solution of linear operators, obtained by integration by parts, usually denoted by the duality bracket $ \big \langle\partial_t \nu_\rightarrow  + v \partial_x \nu_\rightarrow      , \psi  \big \rangle$. Notations $f_{(t^0,x^0)}(dt,dx)$ and $f_{x^e}(dx)$ are used here for what was previously denoted as functions in \eqref{eqint edp flux}. It better conveys the fact that these objects are measures. Of course, when both $f_{(t^0,x^0)}$ and $f_{x^e}$ admit a density with respect to the Lebesgue measure, the weak transport equation are well defined in the classical sense.

Similar to this equation, the flux $\nu_\leftarrow$ from equation \eqref{nu_gauche_simplifie_ti} verifies 
\begin{equation*}
    \partial_t \nu_\leftarrow(t, x)  - v\partial_x \nu_\leftarrow(t, x) 
		        = v F_{x^e}(x) f_{(t^0, x^0)}(t, x) -v \nu_\leftarrow(t, x)\frac{f_{x^e}(x)}{F_{x^e}(x)}. 
\end{equation*}

In the case of a specified arrival time instead of starting time in equations \eqref{nu_droit_simplifie_tf} and \eqref{nu_gauche_simplifie_tf}, we have similarly
\begin{align}
\partial_t \nu_\rightarrow(t, x)  + v\partial_x \nu_\rightarrow(t, x)
&= v\nu_\rightarrow(t, x)\frac{f_{x^0}(x)}{F_{x^0}(x)}   - v F_{x^0}(x) f_{(t^e, x^e)}(t, x), \label{pde_right_te}
\\
\partial_t \nu_\leftarrow(t, x)  - v\partial_x \nu_\leftarrow(t, x) 
&= v \nu_\leftarrow(t, x)\frac{f_{x^0}(x)}{S_{x^0}(x)} - vS_{x^0}(x) f_{(t^e, x^e)}(t, x)\nonumber.
\end{align}	    
Going back to formulas \eqref{nu_right} and \eqref{nu_left}, and their equivalent for arrival times \eqref{nu_right_alternative} and \eqref{nu_left_alternative}, we interpret the oriented flux as a superposition for all different velocities of transport equations with source and death terms.

%%%%%%%%%%%%%
% CONCLUSION
%%%%%%%%%%%%%
\section*{Concluding remarks}

This work introduces a novel framework for the macroscopic modeling of urban active soft mobility, bridging the gap between aggregated, nonintrusive counting data and the fine-grained characterisation of mobility patterns by trip purpose. To ensure interpretability of the activity-based modeling, we derived our quantity of interest from microscopic, stochastic dynamics. A key contribution lies in the rigorous introduction of attendance functions, which encapsulate the spatio-temporal structure of trips associated with different journey types. This critically enables the disaggregation problem to be reformulated as the estimation of subpopulation sizes \textit{via} standard statistical approach. Consistency and asymptotic normality of the estimator is proven, with, in addition, an adapted EM algorithm to offer a practical and efficient computational solution. Beyond its methodological contributions, our framework highlights the potential of sensor-based data as a robust, privacy-preserving alternative to traditional surveys and digital tracking, as one is able to reconstruct meaningful mobility indicators from widely available urban infrastructures.

This article is a first step opening further academic research in that direction. A natural perspective is the extension of the current one-dimensional modeling to realistic two-dimensional urban networks, where trajectory multiplicity, network topology, and multimodality introduce new theoretical and computational challenges. This transition will require coupling local one-dimensional models along network edges with consistent junction dynamics, as well as integrating mode-dependent route choices. We aim to integrate richer data, such as detailed land-use information and synthetic geo-located agendas \cite{colomb2025generation,colomb2026systematic}. This will ensure more realistic attendance functions and thus refine our model. It is also important to notice that it rises significant geovisualization challenges, to account for the joint exploration of mobility intensities and trip purposes of population flows.

Finally, beyond methodological advances, we believe the proposed framework has the potential to become a versatile tool for decision support in sustainable urban mobility. By enabling the reconstruction, interpretation, and comparison of mobility flows by purpose from minimal and non-invasive data, it offers a scalable pathway toward data-informed urban planning, particularly in contexts where traditional or digital mobility data are scarce, biased, or unavailable.

\bibliographystyle{acm}
\bibliography{biblio.bib}

\appendix
\section{Asymptotic study of the maximum likelihood estimator}\label{appendix_stat}

This appendix is dedicated to the theoretical study of the MLE defined in Section~\ref{statistical_model} as the maximum of the likelihood given by formulas \eqref{y_j_likelihood} and \eqref{complete_likelihood}.

\subsection{M-estimation problem}

First, it should be remarked that, if all the $n(i, x_j)$ are zero, then
\begin{equation*}
    \argmax_{\nu \in \RR_+^K}\mathcal{L}\left(\nu, \left(y_j \right)_{j \leq J} \right) = 0_K.
\end{equation*}
Otherwise, if at least one of the $n(i, x_j)$ is nonzero (which is almost surely the case from a certain rank, by virtue of Kolmogorov's zero-one law), then the likelihood is strictly positive for any $\nu\neq 0_K$ and tends to $0$ when $\lVert x \rVert$ tends to infinity or zero. By continuity, the MLE is therefore to be found in a compact set of $\RR_+^K\setminus 0_K$. However, at this point, it is not unique. Straightforward calculations lead to
$$ \widehat{N}_J = \argmax_{\nu \in\RR_+^K \setminus 0_K} 
        \frac{1}{J} \sum_{j=1}^J m(\nu,y_j),$$
where, for any $\nu\neq0_K$,
$$m(\nu, y_j) := \sum_{i=1}^I \left\{-\langle \nu, a(i, x_j) \rangle + n(i, x_j) \log\left(\langle \nu, a(i, x_j) \rangle \right)\right\},$$
which proves in particular that the MLE is an M-estimator. In addition, for $\nu\neq0_K$, $M(\nu)$ denotes the expectation of each term $m(\nu,y_j)$ of the sum,
\begin{align*}
M(\nu) &:= \EE[m(\nu,y)]\\
&= \int_{\mathcal{E}} \sum_{n \in \NN^I} 
       \left( \sum_{i=1}^I \left\{ - \langle \nu, a(i, x) \rangle + n_i\log(\langle \nu, a(i, x) \rangle ) \right\}
        \exp\left(\langle N, a(i, x) \rangle \right) \frac{\langle N, a(i, x) \rangle^{n_i}}{n_i!} \right)f_c(x) dx \\
        &= \sum_{i=1}^I \int_{\mathcal{E}} \left\{ -\langle \nu, a(i, x) \rangle + \langle N, a(i, x_j) \rangle \log(\langle \nu, a(i, x) \rangle \right\} f_c(x)dx.
\end{align*}

\subsection{Uniqueness and consistency of the MLE}

To investigate the asymptotic properties of the MLE, we follow the strategy of \cite[Chapter\, 5]{van_der_vaart} and study the expectation $M(\nu)$. Standard calculations show that
\begin{align}
    \nabla M(\nu) &= \left(\sum_{i=1}^I \int_{\mathcal{E}} a_k(i, x) 
            \left\{ \frac{\langle N , a(i, x) \rangle}{\langle \nu, a(i, x) \rangle} - 1 \right\}
            f_c(x)dx\right)_{k\leq K} , \nonumber\\
     \nabla^2 M(\nu)  &= -\left(\sum_{i=1}^I\int_{\mathcal{E}} a_k(i, x) a_{k'}(i, x) 
            \frac{ \langle N , a(i, x) \rangle}{\langle \nu, a(i, x) \rangle^2 }
            f_c(x) dx \right)_{k, k' \leq K}.         \label{hessienne_M}
\end{align}
The entries of the matrix $\nabla^2 M(\nu)$ can be seen as scalar products of the form $(a_k | a_{k'})$ on the Hilbert space $\LL^2(\llbracket 1, I \rrbracket \times \mathcal{E})$ endowed with
$$\left( F| G \right) := \sum_{i=1}^I \int_{\mathcal{E}}F(i, x)G(i, x) \frac{\langle N, a(i, x) \rangle}{\langle \nu, a(i, x) \rangle^2}f_c(x) dx.$$
It should be noticed that the strict positivity (almost everywhere) of $f_c$ and of attendance functions $a_k$ is essential to ensure that $\langle N, a(i, x) \rangle f_c(x) / \langle \nu, a(i, x) \rangle^2 > 0$ for all $i \leq I$ and $x \in \mathcal{E}$, which in turn guarantees that the inner product $( \cdot | \cdot)$ is positive definite. 

Since the $a_k$'s are assumed to be linearly independent in $\mathcal{C}\left(\llbracket 1, I\rrbracket \times \mathcal{E}\right)$, they are also linearly independent in $\LL^2( \llbracket 1, I \rrbracket \times \mathcal{E})$. Thus, $\nabla^2 M(\nu) = \left( \left( a_k | a_{k'} \right) \right)_{k, k' \leq K}$ is the Gram matrix of an independent family of functions. As such, it is symmetric positive definite for any $\nu \in\RR_+^K\setminus 0_K$, \textit{i.e.} $M$ is strictly concave. In addition, its maximum is reached at $\nu=N$.

Our objective is now to prove that the MLE is uniquely defined from a certain rank, which we address by investigating
$$\widehat{M}_J(\nu) = \frac{1}{J} \sum_{j=1}^J m(\nu,y_j).$$
First, it is obvious to see that
\begin{align*}
    \nabla_{\!\nu} m(\nu, y)
    &= \Bigg(
        \sum_{i=1}^I a_k(i, x)\left\{\frac{n_i}{\langle \nu, a(i, x) \rangle} -1\right\}
        \Bigg)_{k \leq K} ,\\
      \nabla^2_{\!\nu} m(\nu, y)
    &= -\Bigg( 
        \sum_{i=1}^I a_k(i, x) a_{k'}(i, x) \frac{n_i}{\langle \nu, a(i, x) \rangle^2} 
        \Bigg)_{k, k' \leq K}  .
\end{align*}
In particular, the gradient and the Hessian matrix are uniformly continuous in $\nu$ on any compact set $C$ away from $0_K$. Thus, in light of the Glivenko-Cantelli property \cite[Chapter 19]{van_der_vaart}, we have the almost sure uniform convergences over $C$,
\begin{align}
    \sup_{\nu \in C} \norm{\nabla\widehat{M}_J(\nu) - \nabla M(\nu)} &\stackrel{\text{\tiny a.s}}{\longrightarrow} 0 ,\label{eq glivenko cantelli grad} \\
    \sup_{\nu \in C} \norm{\nabla^2\widehat{M}_J(\nu) - \nabla^2 M(\nu)} &\stackrel{\text{\tiny a.s}}{\longrightarrow} 0.\label{eq glivenko cantelli hess}
\end{align}
If $C$ is sufficiently large so that $N \in C$, by continuity and strict concavity of $\nabla^2M$, for any norm $\norm{\cdot}$ over $\mathcal{M}_{K}(\RR)$, there exists a constant $\alpha_1 > 0$ such that
\begin{equation*}
    \sup_{\nu \in C} \norm{\nabla^2 M (\nu)} \geq \alpha_1.
\end{equation*}
Combined with \eqref{eq glivenko cantelli hess}, we deduce that, from a certain rank, almost surely,
\begin{equation*}
    \sup_{\nu \in C} \norm{\nabla^2\widehat{M}_J(\nu)} \geq \frac{\alpha_1}{2},
\end{equation*}
which implies that $\widehat{N}_J$ is almost surely uniquely defined from a certain rank onward.

We now focus on the consistency of the MLE. The strict concavity of $M$ can be rewritten as,
\begin{equation*}\label{concavite gradient}
    \langle \nabla M(\nu), \mu - \nu \rangle > M(\mu) - M(\nu),
\end{equation*}
for any $\nu  \neq \mu$. Using the fact that the unique maximum of $M$ is reached in $N$, it follows that there exists a constant $\alpha_2 > 0$ such that, for all $\nu \in \mathbb{S} (N, \epsilon)$ (the sphere of radius $\epsilon$ centered on $N$), 
\begin{equation*}
    \langle \nabla M(\nu) , N - \nu \rangle > \alpha_2.
\end{equation*}
Let $\epsilon > 0$. We combine this last equation with equation \eqref{eq glivenko cantelli grad} applied on $C= \mathbb{S}(N, \epsilon)$. We deduce that, from a certain rank onward,
\begin{equation*}
    \sup_{\nu \in \mathbb{S}(N, \epsilon)} \langle \nabla\widehat{M}_J(\nu), N-\nu \rangle > \frac{\alpha_2}{2},
\end{equation*}
which states that $\widehat{N}_J$, the maximizer of $\widehat{M}_J$, lies within the ball $\BB(N, \epsilon)$ almost surely, which in turn proves its almost sure convergence toward $N$, by virtue of the Borel-Cantelli lemma.

\subsection{Asymptotic normality and confidence regions}
We will now establish that, as in classical settings, consistency is complemented by asymptotic normality. By leveraging the expression in \eqref{hessienne_M} alongside standard results on Fisher information (see for instance \cite{van_der_vaart}), we obtain a closed-form of the Fisher information matrix \eqref{eq:def:fisher} associated to our statistical model with
$$  \mathcal{I}(N) = \nabla^2 M(N) .$$
As discussed earlier, the Fisher information is in particular invertible, which leads to the following asymptotic normality result (see \cite[Theorem 5.39]{van_der_vaart} for detailed explanations),
\begin{equation*}
    \sqrt{J} \left( \widehat{N}_J - N \right) \stackrel{\text{\tiny d}}{\longrightarrow}\mathcal{N}(0, \mathcal{I}(N)^{-1}).
\end{equation*}
This convergence in distribution enables the construction of confidence regions. By the spectral theorem, we can decompose the Fisher information matrix as $\mathcal{I}(N) = P\Lambda P^\top$, where $P$ is an orthogonal matrix and $\Lambda$ a diagonal matrix with strictly positive entries (since the Fisher information is symmetric positive definite). By stability of convergence in distribution under continuous mapping, we have
$$\sqrt{J}\Lambda^{1/2} P^\top (\widehat{N}_J  -N)\stackrel{\text{\tiny d}}{\longrightarrow}\mathcal{N}(0, I_K).$$
We thus deduce
\begin{equation*}
    J (\widehat{N}_J  -N)^\top
        P \Lambda^{1/2} \Lambda^{1/2} P^\top
        (\widehat{N}_J  -N) 
   \stackrel{\text{\tiny d}}{\longrightarrow} \chi^2_K,
\end{equation*}
where $\chi_K^2$ is the Chi-square distribution with $K$ degrees of freedom, which proves formula \eqref{confidence_interval_ellipsoid} for asymptotic confidence regions for $N$.

\section{Explicit formulas for the EM algorithm}\label{appendix_em}
This appendix is dedicated to the establishment of explicit formulas \eqref{eq:formula:estep} and \eqref{eq:formula:mstep} for the E and M steps of the EM algorithm derived in Section~\ref{section_em}.

\subsection{E step}
In Algorithm~\ref{algo_em}, the E step consists in computing the conditional expectation of the hidden data knowing the count data under any vector of subpopulation sizes $\nu\in\RR_+^K \setminus 0_K$,
 \begin{equation*}
\EE^\text{\tiny h}_\nu\left[ \left. \left( y^h_j \right)_{j \leq J} \right| \left(n(i, x_j)\right)_{i \leq I, j \leq J} \right].
\end{equation*}
Standard calculations yield
 \begin{align*}
 \EE_{\nu}\left[\left( n_k(i, x_j)\right)_{k \leq K} | \ n(i, x_j) = n \right]
 &=\sum_{n_1+\dots+n_K = n}\frac{\displaystyle \PP(n_1(i, x_j)= n_1, \dots, n_K(i, x_j) = n_K)}{\displaystyle \PP(n(i, x_j)=n)} (n_1,\dots,n_K)^\top \\
 &=\frac{1}{\langle\nu, a(i, x_j) \rangle^{n}}\sum_{n_1+\dots+n_K = n}\binom{n}{ n_1, \cdots\!, n_K}\prod_{k=1}^K \left(\nu_k a_k(i, x_j)\right)^{n_k}  (n_1,\dots,n_K)^\top,
 \end{align*}
where $\displaystyle\binom{n}{n_1, \cdots\!, n_K} = \frac{n!}{n_1!\dots n_K!}$. We simplify now each coordinate of the sum, for $q \leq K$,
    \begin{align*}
        & \sum_{ n_1+ \dots +  n_k = n} \binom{n}{n_1, \cdots\!,  n_K} 
            \prod_{k=1}^K \left(\nu_k a_k(i, x_j)\right)^{n_k} n_q \\
        =~& \nu_q a_q(i, x_j) \frac{\partial}{\partial\left( \nu_q a_q(i, x_j) \right)}\left\{\sum_{n_1 + \dots + n_k = n} \binom{n}{n_1, \cdots\!, n_K}
        \prod_{k=1}^K \left(\nu_k a_k(i, x_j)\right)^{n_k} \right\}\\
        =~& \nu_q a_q(i, x_j) \frac{\partial}{\partial \left( \nu_q a_q(i, x_j) \right)} \left\{\sum_{k=1}^K \nu_k a_k(i, x_j) \right\}^{n}\\
        =~& \nu_q a_q(i, x_j)\,n\,\langle \nu, a(i, x_j) \rangle^{n-1}.
    \end{align*}
Consequently,
    \begin{equation*}
    \EE_{\nu}\left[ \left( \left. n_k(i, x_j)\right)_{k \leq K} \right| \ n(i, x_j)  \right] =  \left(n(i, x_j) \frac{\nu_k a_k(i, x_j)}{\langle\nu, a(i, x_j) \rangle} \right)_{k \leq K}.
    \end{equation*}
    Since $n_k(i, x_j)$ does not depend on $n(i', x_{j'})$ if $i'\neq i$ or $j' \neq j$, this states formula \eqref{eq:formula:estep} for the E step of Algorithm~\ref{algo_em}.

\subsection{M step}
The M step of Algorithm~\ref{algo_em} consists in finding the maximum of the hidden likelihood for any hidden data $(z_{k,i,j})_{k \leq K, i \leq I, j \leq J}$, 
\begin{equation*}
    \argmax_{\nu \in \RR_+^K} \mathcal{L}^{\text{\tiny h}} \left(\nu, \left(x_j, \left(z_{k,i,j} \right)_{k \leq K, i \leq I} \right)_{j \leq J} \right) = \argmax_{\nu \in \RR_+^K} \prod_{j=1}^J   \mathcal{L}^{\text{\tiny h}} \left(\nu, \left(x_j, \left(z_{k,i,j} \right)_{k \leq K, i \leq I} \right) \right).
\end{equation*}
 In the formula \eqref{likelihood_hidden_data} of the likelihood with hidden data, for $k \leq K$, $\nu_k$ appears only in the $k$-th factor of $\mathcal{L}^{\text{\tiny h}}$. Maximizing $\mathcal{L}^{\text{\tiny h}}$ is therefore equivalent to maximizing each of the bracketed factors. For any $(y^{\text{\tiny h}}_j)_{j\leq J} =  (x_j, (n_k(i, x_j))_{i\leq I, \ k \leq K})_{j \leq J}$, we seek
\begin{equation*}
    \argmax_{\nu_k \in \RR_+}
        \prod_{j=1}^J\left\{
            f_c(x_j)\prod_{i=1}^I  
                \exp\left(-a_k(i, x_j)\nu_k\right) \frac{\left(a_k(i, x_j)\nu_k\right)^{n_k(i, x_j)}}{n_k(i, x_j)!}
        \right\}.
\end{equation*}
If all the $n_k(i, x_j)$'s are zero, it is easy to see that the argument of the maximum is $0$. Otherwise, we seek an argument of the maximum in $\RR_+^\ast$, where the log-likelihood is well-defined. By the monotonicity of the logarithm, straightforward calculations show that
        \begin{align*}
        & \argmax_{\nu_k \in \RR_+^\ast}
            \prod_{j=1}^J\left\{
                f_c(x_j)\prod_{i=1}^I  
                    \exp\left(-a_k(i, x_j)\nu_k\right)\frac{\left(a_k(i, x_j)\nu_k\right)^{n_k(i, x_j)}}{n_k(i, x_j)!}
            \right\} \\
        =~ & \argmax_{\nu_k \in \RR_+^\ast}\sum_{i=1}^I \sum_{j=1}^J 
                -a_k(i, x_j)\nu_k + n_k(i, x_j)\log \left( \nu_k \right) \\
        =~ & \argmax_{\nu_k \in \RR_+^\ast} \ 
                \log(\nu_k)\frac{\sum_{i, j}n_k(i, x_j)}{\sum_{i, j} a_k(i, x_j)}  - \nu_k  \\
        =~ & \frac{\sum_{i, j}n_k(i, x_j)}{\sum_{i, j} a_k(i, x_j)}.
    \end{align*}
 Interestingly, the formula also applies if all the $n_k(i, x_j)$'s are zero. Finally, this proves formula \eqref{eq:formula:mstep} of Algorithm~\ref{algo_em}.

\section{Finding the attendance function and transport equations}\label{appendix_attendance}

This appendix presents the technical developments used in Section~\ref{stochastic_deterministic} to derive the attendance function within the framework of transport equations.

\subsection{Expression of the attendance function as the integral of a flux}\label{subsec app att1}

We explain here the link between the attendance $a_k(i, x)$ and  the flux that we introduced in formula \eqref{eq ak flux}. More precisely, we focus on the  case of an arrival time distribution, showcased  in Subsection~\ref{alternative_model}. We rigorously derive formula \eqref{eq attendance arrival time} with explicit expressions of the flux \eqref{nu_right_alternative} and \eqref{nu_left_alternative}, in a probabilistic framework, and in the limit of many trips.

This proof is still valid, up to some changes of signs and notations, in the in case of a starting time distribution used in Subsections~\ref{single_individual} and \ref{large_population}. The same procedure works to establish formulas \eqref{nu_left} and \eqref{nu_right}, as the only difference lies in the expression of the trajectory $\gamma$: \eqref{def traj} can be compared with \eqref{gamma_arrival_time}. For clarity reasons, as the journey type $k$ is fixed at the beginning of the modeling of the attendance in Subsection~\ref{single_individual}, we drop altogether the index $k$.

We draw $( {v},  {x^0},  {t^e},  {x^e}, \gamma, \mathcal{T})$ a family representing the velocity, starting point, ending time, ending point trajectory, and travel interval of an individual with journey type $k$, according to the distribution of the  random variables $(v_p, x^0_p, t^e_p, x^e_p, \gamma_p, \mathcal{T}_p)_{p \leq N_k}$ defined in Subsection~\ref{single_individual}. In particular, we have shown in Subsection~\ref{single_individual} the seminal relationship
\begin{align*}
    a(i, x) = \PP\big(x \in \gamma(T_i \cap \mathcal{T})\big ).
\end{align*}
It stands for  the probability that the position of the counter, $x$, is along the trajectory of the individual, $\gamma$, during the counting time step  $T_i$, and also the duration of its trip, $\mathcal T$. We now compute this probability. 

In a one-dimensional setting, one can decompose $a(i, x)$ with rightward and leftward contributions. This amounts to the choice of a destination point, given a starting position. Considering the disjoint events $\{{x^0} \leq  {x^e}\}$ and $\{{x^e} <  {x^0}\}$, we have
\begin{multline*}
    \PP \big(x \in \gamma(T_i \cap \mathcal{T}) \big) \\
 = \PP \Big( \left\{\gamma^{-1}(x) \in T_i \right\} \cap\left\{ x \in \gamma(\mathcal{T}) \right\} \cap \left\{{x^0} \leq  {x^e} \right\}   \Big)
	 + \PP \Big( \left\{\gamma^{-1}(x) \in T_i \right\} \cap\left\{ x \in \gamma(\mathcal{T}) \right\} \cap \left\{ {x^e} < {x^0} \right\} \Big). 
\end{multline*}
Depending on the directions of travel, we plug in different explicit expressions for $\gamma$, according to \eqref{gamma_arrival_time}. We find
\begin{align*}  
 \PP \big(x \in \gamma(T_i \cap \mathcal{T}) \big) & =  \PP \left(\left\{  {t^e} - \frac{ {x^e}-x}{ {v}} \in T_i\right\}
		\cap \left\{ {x^0} \leq x <  {x^e} \right\}  \right) 
		+ \PP \left( \left\{  {t^e} - \frac{x- {x^e}}{ {v}} \in T_i\right\}
		\cap \left\{ {x^e}  < x \leq  {x^0} \right\}  \right)	.
 \end{align*}
By independence of $ {x^0}$ with other random variables, we have further, by definition of the cumulative distribution function $F_{x^0}$ and of the survival function $S_{x^0}$ of the starting point,    
\begin{multline*}
	\PP \big(x \in \gamma(T_i \cap \mathcal{T}) \big) = 
 		F_{x^0}(x) \, 		\PP \left( \left\{  {t^e} -\frac { {x^e}-x}{ {v}} \in T_i \right\}
		\cap \left\{ {x^e}-x \geq 0 \right\} \right)  \\
		+ S_{x^0}(x) \,	\PP \left( \left\{ {t^e} - \frac{x- {x^e}}{ {v}} \in T_i \right\}
		\cap \left\{x -  {x^e} \geq 0  \right\}\right).
 \end{multline*}
In this last expression, we have expressed the probability $a(i, x)$ as the sum of two terms: the probability that an individual arrives at $x$ from the left, \textit{i.e.} when $ {x^0}$ is to the left of $x$, and the probability that an individual arrives at $x$ from the right, \textit{i.e.} when $ {x^0}$ is to the right of $x$. The probability that an individual arrives at $x$ from the left is then the product of the probability than $x_0$ is to the left of $x$, 	$F_{x^0}(x)$, and the probability that the individual arrives at $x$ during the adequate time period. In this last probability, the term $t^e - (x^e-x)/v$ is  the time remaining (before reaching the position $x^e$) in the individual's trip, when it is at position $x$ and travels at speed $v$. 

We will now express these probabilities, using the generic integral form. To do so, we use the conditional density of $ {t^e}$ knowing $ {x^e}=e$, leveraging the independence between $( {t^e},  {x^e})$ and ${v}$,
\begin{multline*}
    a(i,x) = F_{x^0}(x) \int_{\RR_+^\ast} \left\{  \int_{\RR}   \int_{\RR} 
		\ind_{\left\{t - \frac{e-x}v  \in T_i \right\}} \ind_{\left\{(e-x) \geq 0  \right\}} f_{t^e|x^e}(t|e) dt\,f_{x^e}(de)	\right\}  f_v(dv)  
		+ \\ 
		S_{x^0}(x) \int_{\RR_+^\ast} \left\{ \int_{\RR}  \int_{\RR} 
		\ind_{\left\{ t - \frac{e-x}v  \in T_i \right\}}  \ind_{\left\{x-e\geq 0\right\}} f_{t^e|x^e}(t|e) dt\,f_{x^e}(de) \right\}f_v(dv).
\end{multline*}
Thanks to the indicator function, we can change the integrands in $de$,
\begin{multline*}
	 a(i,x) =	F_{x^0}(x) \int_{\RR_+^\ast} \left\{  \int_{x}^{+\infty}  \int_{ \RR }
	\ind_{\left\{t - \frac{e-x}v  \in T_i \right\}}   f_{t^e|x^e}(t | e) 
		dt \, f_{x^e}(de) \right\} f_v(dv)  	+ \\
	 S_{x^0}(x) \int_{\RR_+^\ast} \left\{  \int_{-\infty}^{x}   \int_{ \RR } 
		\ind_{\left\{t - \frac{e-x}v  \in T_i \right\}} f_{t^e|x^e}(t|e) 
		dt \, f_{x^e}(de) \right\} f_v(dv).
\end{multline*} 
For the last indicator function, we apply the change of variable $t$ instead of $t - (e-x)/v$,
\begin{multline*}
 a(i,x) =		F_{x^0}(x) \int_{\RR_+^\ast} \left\{ \int_x^{+\infty} 
	  \int_{T_i} f_{t^e|x^e} \left(t + \left. \frac{e-x}{v} \right| e \right) dt \,
	 f_{x^e}(de) 	\right\} f_v(dv)+ \\
		 S_{x^0}(x) \int_{\RR^\ast_+} \left\{  \int_{-\infty}^{x} 
	 \int_{T_i} f_{t^e|x^e} \left( t + \left. \frac{x-e}{v} \right| e \right) dt \,
	 f_{x^e}(de) 	\right \}f_v(dv).
	\end{multline*}
We can now apply Fubini's theorem,
\begin{multline*}
	a(i, x) = 
		F_{x^0}(x)\int_{T_i} \int_x^{+\infty} 
		\int_{\RR^\ast_+} f_{t^e|x^e} \left(t + \left. \frac{e-x}{v} \right|  e \right) f_v(dv)
		f_{x^e}(de)\,dt + 
		\\
		 S_{x^0}(x)\int_{T_i} \int_{-\infty}^{x} 
		\int_{\RR^\ast_+} f_{t^e|x^e}  \left( t + \left. \frac{x-e}{v} \right| e \right)
		f_v(dv)f_{x^e}(de)\,dt,
	\end{multline*}
which gives the form, as in \eqref{eq attendance arrival time},
\begin{equation*}
    a(i, x) 
    = \int_{T_i} \nu_\rightarrow(t, x) 
    +  \int_{T_i}\nu_\leftarrow(t, x),
\end{equation*}
with the exact formulas \eqref{nu_right_alternative} and \eqref{nu_left_alternative}. Equivalently, if we want a more compact form, we can write from the last equation,
\begin{equation*}
    a(i, x) = \int_{T_i} \int_{\RR_+^\ast} \int_\RR G_{x}(u) f_{t^e|x^e} \left( \left. t+ \frac{|u-x|}{v} \right| u \right), 
\end{equation*}
where $G_x$ is defined in \eqref{eq:def:Gx}.

\subsection{Derivation of transport equations}\label{subsec app att2}
We provide here a  detailed derivation of the transport equation \eqref{pde_right_te} from  the rightward flux of an ending time distribution, as defined in \eqref{nu_droit_simplifie_tf}. The other cases follow analogously.

\noindent \textbf{Step 1.} Let $\psi \in \mathscr{C}_c^{\infty}(\RR\times\RR)$ a test function. We will integrate the flux against $\psi$ to find the PDE.  For $y \in \mathcal{E}$ a spatial position, we define the trajectory that moves  at speed $v \neq 0$, and  is at the position $y$ at time $0$: $ t \in \RR \mapsto y + vt$.  With the affine change of variable $x = y + vt $,
\begin{multline*}
    \Big \langle \partial_t \nu_\rightarrow + v\partial_x \nu_\rightarrow , \psi \Big \rangle_{\RR \times \RR} 
      := \int_\RR \int_\RR \nu_\rightarrow(t, x) 
        \Big(-\partial_t \psi(t, x) - v \partial_x \psi(t, x) \Big) dx\,dt \nonumber \\
     = \int_\RR \int_\RR \nu_\rightarrow (t, y + vt) 
        \Big(-\partial_t \psi(t, y + vt) - v \partial_x \psi (t, y + vt) \Big) 
        dy\,dt .
        \end{multline*}
With the additional notation $\Gamma_y(t) = (t, y + vt)$, we have
        \begin{align}
          \Big \langle \partial_t \nu_\rightarrow + v\partial_x \nu_\rightarrow , \psi \Big \rangle  =    \int_\RR  \int_\RR -\nu_\rightarrow \circ \Gamma_y(t) (\psi \circ \Gamma_y)'(t) dt\,dy      = \int_\RR \Big \langle \left(\nu_\rightarrow \circ \Gamma_y \right)', \psi \circ \Gamma_y \Big \rangle_{\RR} \,  dy.\label{step_1_final}
        \end{align}
        
\textbf{Step 2.} Let $x \in \RR$ and $\varphi \in \mathscr{C}_c^\infty (\RR )$ another test function. We introduce the notation $g_x$ for the affine trajectory  $g_x(t) = x  + vt$. We have
\begin{align*}
\Big 	\langle ( \nu_\rightarrow \circ \Gamma_x )' , \varphi \Big \rangle_\RR 
	&= \Big \langle \nu_\rightarrow \circ \Gamma_x , -\varphi' \Big \rangle_\RR = \int_\RR -\nu_\rightarrow \big (t, g_x(t) \big) \varphi'(t) dt, \\
	&= \int_\RR -F_{x^0}\big(g_x(t)\big) \int_{g_x(t)}^{+\infty} f_{t^e|x^e} \left( \left. \frac{u-x}{v} \right| u \right) f_{x^e}(du) \varphi'(t)dt,
\end{align*}
where we have used the definition of the flux \eqref{nu_droit_simplifie_tf} for the last equality. Since $g_x$ is invertible for all $x \in \mathcal{E}$, with inverse given by $g_x^{-1}(u) = (u-x)/v$ for all $u \in \RR$, then $u \geq g_x(t)$ if and only if $t \leq g_x^{-1}(u)$. We therefore obtain, by applying Fubini's theorem to our last expression,
\begin{align}\label{transformation_1}
\Big 	\langle (\nu_\rightarrow \circ \Gamma_x )' , \varphi \Big \rangle_\RR 
   &= \int_\RR \left\{ \int_{-\infty}^{g_x^{-1}(u)} -F_{x^0}\big(g_x(t)\big) \varphi'(t) dt \right\} f_{t^e|x^e} \left( \left. g_x^{-1}(u) \right| u \right) f_{x^e}(du).
\end{align}		
The inner integral can be rewritten, thanks to the affine change of variable $t = g_x^{-1}(a)$, 
\begin{align*}
\int_{-\infty}^{g_x^{-1}(u)} -F_{x^0}\big(g_x(t)\big) \varphi'(t) dt   = \int_{-\infty}^u -F_{x^0}(a) \left( \varphi \circ g_x^{-1} \right)'(a) da.
	\end{align*}
Then we can do an integration by part, in the Stieltjes sense,
\begin{multline}
 \int_{-\infty}^{g_x^{-1}(u)} -F_{x^0}\big(g_x(t)\big) \varphi'(t) dt 
	 = \left[ -(\varphi \circ g_x^{-1}) F_{x^0} \right]_{-\infty}^u 
	    + \int_{-\infty}^u \varphi(g_x^{-1}(a)) f_{x^0}(da),   \\
     = -\varphi\big( g_x^{-1}(u)\big) F_{x^0}(u) 
	    + \int_{-\infty}^u \varphi \big(g_x^{-1}(a)\big) f_{x^0}(da)\label{integration_by_part}.
\end{multline}	
We finally plug \eqref{integration_by_part} into \eqref{transformation_1} to deduce
\begin{multline*}
\Big	\langle (\nu_\rightarrow \circ \Gamma_x )' | \varphi \Big \rangle_\RR 
 =  \int_\RR -\varphi\big(g_x^{-1}(u)\big) F_{x^0}(u)
		f_{t^e | x^e} \left( \left.  g_x^{-1}(u) \right| u  \right) f_{x^e} (du) \\
		+ \int_{\RR} \left\{\int_{-\infty}^u 
		\varphi( g_x^{-1}(a)) f_{x^0}(da) \right\}
		f_{t^e | x^e} \left( \left. g_x^{-1}(u) \right| u \right) f_{x^e} (du) .
\end{multline*}
With another application of Fubini's theorem on the second term,
\begin{multline*}
\Big	\langle (\nu_\rightarrow \circ \Gamma_x )' | \varphi \Big \rangle_\RR 
 =  		\int_\RR -\varphi\big(g_x^{-1}(u)\big) F_{x^0}(u)
		f_{t^e | x^e} \left( \left.  g_x^{-1}(u) \right| u  \right) f_{x^e} (du)  \\
		+ \int_{\RR} \left\{\int_a^{+\infty} 
		        f_{t^e | x^e} \left( \left. g_x^{-1}(u) \right| u \right)
		        f_{x^e} (du) \right\} \varphi\big( g_x^{-1}(a) \big) f_{x^0}(da) .
	\end{multline*}
We recognize the definition of the flux \eqref{nu_droit_simplifie_tf} in this last term,
 \begin{multline}\label{step_2_final}
	\Big	\langle (\nu_\rightarrow \circ \Gamma_x )' | \varphi \Big \rangle _\RR
 =  	\int_\RR -\varphi\big(g_x^{-1}(u)\big) F_{x^0}(u)
		f_{t^e | x^e} \left( \left.  g_x^{-1}(u) \right| u  \right) f_{x^e} (du)  \\
		+ \int_{\RR} \frac{1}{F_{x^0}(a)} \nu_\rightarrow(g_y^{-1}(a), a) 
		    \varphi\big( g_x^{-1}(a)\big) f_{x^0}(da) .
	\end{multline}

\textbf{Step 3.} We now plug \eqref{step_2_final} into \eqref{step_1_final} with $\varphi = \psi \circ \Gamma_y$,  to obtain
\begin{multline*}
   \Big \langle \partial_t \nu_\rightarrow + v \partial_x \nu_\rightarrow, \psi \Big \rangle_{\RR \times \RR}
    =
			\int_\RR  \int_\RR -\psi(g_y^{-1}(u), u) F_{x^0}(u)
			f_{t^e | x^e} \left( \left.  g_y^{-1}(u) \right| u  \right) f_{x^e} (du)\, dy \\
			+ \int_\RR  \int_{\RR} \frac{1}{F_{x^0}(a)} \nu_\rightarrow(g_y^{-1}(a), a) \psi( g_y^{-1}(a), a) f_{x^0}(da)\,dy . 
		\end{multline*}
				With the explicit expression of $g_y^{-1}(u)$,
\begin{multline*}
	   \Big \langle \partial_t \nu_\rightarrow + v \partial_x \nu_\rightarrow, \psi \Big \rangle_{\RR \times \RR} = 		\int_\RR \int_\RR -\psi \left( \frac{u-y}{v}, u \right) F_{x^0}(u)
			f_{t^e | x^e} \left(   \left. \frac{u-y}{v} \right| u  \right) dy\,f_{x^e} (du) \\
			  + \int_\RR   \int_{\RR} \frac{1}{F_{x^0}(a)} \nu_\rightarrow \left( \frac{a-y}{v} , a \right) \psi \left( \frac{a-y}{v}, a \right) dy\,f_{x^0}(da).
		\end{multline*}
We conclude with another affine change of variable,
\begin{multline*}
   \Big \langle \partial_t \nu_\rightarrow + v \partial_x \nu_\rightarrow, \psi \Big \rangle_{\RR \times \RR}  
     =  
			\int_\RR \int_\RR -\psi \left( t, u \right) F_{x^0}(u)
			f_{t^e | x^e} ( t | u ) v\,dt\,f_{x^e} (du) \\
			+ \int_\RR \int_{\RR} \frac{1}{F_{x^0}(a)} \nu_\rightarrow \left( t , a \right) \psi \left( t , a \right) v\,dt\,f_{x^0}(da).
		\end{multline*}
By definition of $f_{t^e|x^e}$, we recognize the distribution $f_{(t^e, x^e)}(t, u) =  f_{t^e|x^e}(t|u)f_{x^e}(u)$,
		\begin{multline*}
		  \Big \langle \partial_t \nu_\rightarrow + v \partial_x \nu_\rightarrow, \psi \Big \rangle_{\RR \times \RR} = 	     \int_{\RR^2} - \psi(t, u) v F_{x^0}(u) f_{(t^e, x^e)}(dt, du) 
	    + \int_\RR \int_\RR \psi (t, a) v \nu_\rightarrow (t, a) \frac{1}{F_{x^0}(a)} dt f_{x^0}(da)  \\ 
 = \Big \langle -v F_{x^0}(x) f_{(t^e, x^e)}(dt, dx) + v \nu_\rightarrow(t, x) \frac{1}{F_{x^0}(x)}f_{x^0}(dx), \psi \Big \rangle_{\RR \times \RR}. 
		\end{multline*}
In other words,
\begin{equation*}
    \partial_t \nu_\rightarrow(t, x) + v \partial_x \nu_\rightarrow(t, x) = -v F_{x^0}(x) f_{(t^e, x^e)}(t, x) + v \nu_\rightarrow(t, x) \frac{f_{x^0}(x)}{F_{x^0}(x)},
\end{equation*}
in the weak sense, on the dual of the space of test functions of $\mathscr{C}_c^\infty(\RR\times\RR)$, which is what we claimed in \eqref{pde_right_te}.

\end{document}